\documentclass[useAMS,usenatbib,usegraphicx]{mn2e}

\usepackage{amssymb} 
\usepackage{textcomp} 
\usepackage{appendix}
\usepackage{amsmath} 
\usepackage{fixltx2e} 
\usepackage{multirow} 
\usepackage{enumerate} 
\usepackage{mathptmx} 
\usepackage{lscape}  
\usepackage{hyperref} 

\newcommand{\kms}{\rm km\ s^{-1}}
\newcommand{\ergs}{\rm erg\ s^{-1}}

\newcommand{\kev}{\rm keV}

\let\AAold\AA
\renewcommand{\AA}{\text{\AAold}}

\newcommand{\lbol}{L_{\rm bol}}
\newcommand{\lledd}{L / L_{\rm{Edd}}}
\newcommand{\mbh}{M_{\rm BH}}

\newcommand{\aox}{\alpha_{\rm ox}}

\newcommand{\nln}{\nu L_{\nu}}

\newcommand{\msun}{{\rm M_{\odot}}}

\newcommand{\hii}{\text{H~{\sc ii}}}
\newcommand{\nii}{\text{[N~{\sc ii}]}}
\newcommand{\sii}{\text{[S~{\sc ii}]}}

\newcommand{\oi}{\text{[O~{\sc i}]}}
\newcommand{\oii}{\text{[O~{\sc ii}]}}
\newcommand{\oiii}{\text{[O~{\sc iii}]}}

\newcommand{\feii}{\text{Fe~{\sc ii}}}
\newcommand{\caii}{\text{[Ca~{\sc ii}]}}
\newcommand{\neiii}{\text{[Ne~{\sc iii}]}}
\newcommand{\nev}{\text{[Ne~{\sc v}]}}

\newcommand{\Ha}{\text{H$\alpha$}}

\newcommand{\Hb}{\text{H$\beta$}}

\newcommand{\aap}{A\&A}
\newcommand{\araa}{ARA\&A}
\newcommand{\apjl}{ApJ}
\newcommand{\apjs}{ApJS}
\newcommand{\apj}{ApJ}
\newcommand{\aj}{AJ}
\newcommand{\mnras}{MNRAS}
\newcommand{\lbha}{L_{\rmn{bH\alpha}}}
\newcommand{\lcont}{L_{\rm cont}}
\newcommand{\lbhb}{L_{\rmn{bH\beta}}}

\newcommand{\lnha}{L_{\rmn{nH\alpha}}}
\newcommand{\lnhb}{L_{\rmn{nH\beta}}}

\newcommand{\loiii}{L_{\oiii}}
\newcommand{\loi}{L_{\oi}}
\newcommand{\lsii}{L_{\sii}}
\newcommand{\lnii}{L_{\nii}}

\newcommand{\lx}{L_{\rm X}}
\newcommand{\luv}{L_{\rm UV}}
\newcommand{\ewha}{\rm{EW}_{\rmn{bH\alpha}}}

\newcommand{\dv}{\Delta {\rm v}}

\newcommand{\lagn}{L_{\rm AGN}}
\newcommand{\lhost}{L_{*}}

\newcommand{\cfnlr}{{\rm CF}_{\rm NLR}}
\newcommand{\znlr}{Z_{\rm NLR}}

\newcommand{\zblr}{Z_{\rm BLR}}

\newcommand{\ncrit}{n_{\rm crit}}

\newcommand{\nd}{$^{\rm nd}$}
\newcommand{\rd}{$^{\rm rd}$}
\renewcommand{\th}{$^{\rm th}$}

\title{Type 1 AGN at low $z$. III. The optical narrow line ratios}
\author[Jonathan Stern and Ari Laor]
{Jonathan Stern\thanks{E-mail: \href{mailto:stern@physics.technion.ac.il}{stern@physics.technion.ac.il} (JS);\newline \href{mailto:laor@physics.technion.ac.il}{laor@physics.technion.ac.il} (AL)}
and Ari Laor\footnotemark[1]\\
Department of Physics, Technion -- Israel Institute of Technology, Haifa 32000, Israel}

\begin{document}
\maketitle

\begin{abstract}
We present the optical narrow line ratios in an SDSS based sample of 3\,175
broad \Ha\ selected type 1 AGN, and explore their positions in the BPT
diagrams as a function of the AGN and the host properties.
We find the following:
1. The luminosities of all measured narrow lines (\Ha,\Hb,\oiii,\nii,\sii,\oi)
show a Baldwin relation relative to the broad \Ha\ luminosity $\lbha$, with
slopes in the range of $0.53-0.72$.
2. About 20\% of the type 1 AGN reside within the `Composite' and `SF' regions
of the BPT diagrams.
These objects also show excess narrow \Ha\ and UV luminosities, for their
$\lbha$, consistent
with contribution from star formation which dominates the narrow lines
emission, as expected
from their positions in the BPT diagrams.
3. The type 1 which reside within the AGN region in the BPT diagrams, are offset to
lower \sii/\Ha\ and \nii/\Ha\ luminosity ratios, compared to type 2 AGN.
This offset is a selection effect, related to the lower AGN/host luminosity
selection of the type
2 AGN selected from the SDSS galaxy sample.
4. The \nii/\Ha\ and \nii/\sii\ ratios in type 1 AGN increase with the host
mass, as expected if the mass-metallicity relation of quiescent galaxies
holds for the AGN narrow line region.
5. The broad lines optical \feii\ is higher for a higher  \nii/\Ha, at a
fixed $\lbol$ and Eddington ratio $\lledd$. This suggests that the broad line
region metallicity is also related to the host mass.
6. The fraction of AGN which are LINERs increases sharply with decreasing
$\lledd$.
This fraction is the same for type 1 and type 2 AGN.
7. The BPT position is unaffected by the amount of dust extinction of the
optical-UV continuum, which suggests the extincting dust resides on scales
larger than the NLR.
\end{abstract}


\section{Introduction}

The gas located on 1 -- 1\,000 pc scale from the center of Active Galactic Nuclei (AGN) plays a role in several important processes, which are not well understood.
This gas is the source of AGN fuel, and may absorb AGN energy and momentum output, thus potentially coupling the growth of the bulge with the growth of the central black hole. 
It is enriched during the life cycle of stars near the nucleus, and therefore traces the star formation history. It also reprocesses the AGN ionization continuum, which originates from a few Schwarzschild radii, and thus its emission may allow to constrain the accretion mode in the innermost regions. 

The most prominent optical signature of the circumnuclear gas in AGN is its emission lines, which have widths typical of the galaxy potential ($\sim 300\ \kms$). These lines are known as the narrow emission lines, and the emitting region as the narrow line region (NLR). 
The vast majority of NLR analyses were performed on type 2 AGN where the central source is obscured, 
partly because the narrow lines are not blended with the broad emission lines, which dominate
the emission features in unobscured type 1 AGN. Most previous studies of the NLR of type 1 AGN were either limited to the most prominent forbidden lines (e.g. Boroson \& Green 1992, using \oiii\ $\lambda5007$), limited to small samples (e.g. Baldwin, Phillips \& Terlevich 1981, hereafter BPT, Cohen 1983, Ho et al. 1997b, Rodr{\'{\i}}guez-Ardila et al. 2000, V{\'e}ron-Cetty et al. 2001, Dietrich et al. 2005), or limited to samples of very weak type 1 AGN (e.g. Greene \& Ho 2007) in which the narrow lines become more prominent (Stern \& Laor 2012b, hereafter Paper II). 

A measurement of narrow line luminosities of a large sample of type 1 AGN, including luminous quasars, was performed by Zhang et al. (2008). They found that the narrow line luminosity ratio $\nii \lambda 6583 / \Ha$ of type 1 AGN is offset to lower values than in type 2 AGN. 
Here we significantly expand their work, by studying the NLR properties of a large sample of 3\,175 type 1 AGN,
hereafter the T1 sample, defined in Stern \& Laor (2012a, hereafter Paper I) with minor adjustments detailed below. The T1 sample spans a black hole mass range of $10^{6}<\mbh<10^{9.5}\ \msun$ and a bolometric luminosity range of $10^{42}<\lbol<10^{46}\ \ergs$. 
In contrast with studies of type 2 AGN, here the AGN is unobscured. We use the narrow line measurements, combined with the AGN spectral energy distribution (SED) and broad line measurements, to address the following questions:

{\it How complete is the BPT classification of AGN?}
The BPT diagrams (BPT and Veilleux and Osterbrock 1987, hereafter VO) compare the ratio of the \oiii\ to \Hb\ luminosity (for brevity \oiii/\Hb), with $\nii \lambda 6583/\Ha$, $\sii(\lambda\lambda 6716, 6731)/\Ha$, and $\oi(\lambda 6300)/\Ha$. These line ratios provide a measure of the relative strength
of the higher energy ionizing photons, and thus differentiate between stellar and AGN excitation. These diagrams are widely used to define type 2 AGN samples, using separation lines based on theoretical models (Kewley et al. 2001, hereafter Ke01), and based on the observed distribution of star forming galaxies (Kauffmann et al. 2003, hereafter Ka03). 

The BPT/VO AGN selection criteria are commonly viewed as necessary and sufficient conditions to define AGN.
However, AGN samples selected by other means show these selection criteria may not be necessary conditions.
In a hard X-ray selected sample, a unique signature for AGN emission, Winter et al. (2010) found that five out of 60 objects are in the Star Forming (SFs) galaxies regime, i.e. below the Ka03 line in the \nii/\Ha\ panel of the BPT diagrams, and five more are between the Ka03 line and the Ke01 line, i.e. `Composites'. In the $\mbh < 10^{6.2}\ \msun$ type 1 sample of Greene \& Ho (2007), 39\% of the objects are SFs or Composites. This fraction dropped to 18\% when the spectra was taken from a narrower slit (Xiao et al. 2011). On the other hand, only 3\% of radio loud AGN are classified as Composites or SFs (Buttiglione et al. 2010).
Using the T1 sample, which is selected independently of the narrow line properties, we derive the completeness of the BPT-based selection criteria, and its dependence on the AGN emission properties.

{\it How are the properties of the NLR gas related to AGN and host properties?}
In low $z$ type 2 AGN, the value of \nii/\Ha, which follows NLR metallicity, $\znlr$, has been found to modestly increase with host mass $M_*$ (Groves et al. 2006) and with host velocity dispersion $\sigma_*$ (Annibali et al. 2010). These trends are associated with the known $M_* - Z$ relation of quiescent galaxies (Lequeux et al. 1979, and citations thereafter). The $\znlr-M_*$ relation is also implied by the fact that most AGN reside in massive galaxies (Ka03) and have $\znlr>Z_\odot$ (Storchi-Bergmann et al. 1998, Groves et al. 2004, 2006), while the rare low $M_*$ AGN have low $\znlr$ (Kraemer et al. 1999, Barth et al. 2008; Ludwig et al. 2012). 
However, these samples are dominated by low $\lbol$ AGN, since they are based on the detectability of the host galaxy, and therefore are limited to a small volume where luminous AGN are rare. 

In high $\lbol$ AGN at high $z$, an $M_* - Z$ relation can be inferred from the increase of $\zblr$ with $\lbol$ (Hamman \& Ferland 1993, 1999, Nagao et al. 2006a), and a likely relation of $\lbol- M_*$.
Though $\znlr$ and $\zblr$ are related (Shields et al. 2010), there seems to be another variable beyond $M_*$ which determines $\zblr$, probably related to the accretion rate in Eddington units ($\lledd$, Shemmer \& Netzer 2002, Shemmer et al. 2004, Shields et al. 2010). Therefore, it is interesting to compare $\znlr$ with $\lbol$ directly. 
Most narrow line measurements in high $\lbol$ AGN are based on narrow line radio galaxies samples (De Breuck et al. 2000, Vernet et al. 2001, Iwamuro et al. 2003, Nagao et al. 2006b). These studies measured UV line ratios, except Iwamuro et al. which measured non-BPT optical line ratios. Comparison of NLR properties derived from different lines can be ambiguous, due to degeneracies in the photoionization models (Nagao et al. 2006b). 
Therefore, the dependence of $\znlr$ and other NLR properties on $\lbol$ is still an open question. In this work we derive indicators of $\znlr$ based on the BPT ratios, for a large dynamical range in $\lbol$. Using the large size of the T1 sample, we also decouple the dependence of $\znlr$ on $\lbol$ and on $M_*$, and compare $\znlr$ with $\zblr$. 

{\it Is the ratio of UV to X ray luminosity a measure of the slope of the ionizing spectrum?}
Due to Galactic absorption, the ionizing part of the AGN spectrum in the extreme UV is generally unavailable. 
Laor et al. (1997) showed that the mean 2~\kev\ luminosity $\lx$ of PG quasars is consistent with an extrapolation of the mean EUV slope (Zheng et al. 1997, Telfer et al. 2002). Therefore, the interpolated slope between $\luv$ and $\lx$, $\aox$, may provide a good estimate of the true ionizing slope. Since the BPT diagrams provide an independent constraint on the ionizing slope, we explore this hypothesis by comparing the BPT ratios with $\aox$ in the T1 sample.

A related issue concerns the location of the optically thin dust found in type 1 AGN samples (Richards et al. 2003, Gaskell et al. 2004, Paper I), which can harden $\aox$. If the extincting dust is located within the NLR, the NLR will see a harder spectrum, and the BPT ratios are expected to vary with the amount of reddening. 
If the extincting dust resides outside the NLR, the NLR will absorb the original ionizing spectrum, and the BPT ratios will remain constant. 
Below, we constrain the location of the extincting dust using the BPT diagrams. 

{\it Is the Seyfert-LINER transition related to other emission properties?}
Kewley et al. (2006, hereafter Ke06) found a bimodality in the BPT diagrams between high ionization Seyferts and low ionization nuclear emission line regions (LINERs, Heckman 1980). They showed the Seyfert-LINER transition is related to $\lledd$, as noted previously by Ho (2002). This transition has also been claimed to be related to the existence of the broad lines, due to the low detection fraction of broad lines in LINERs (Ho et al. 1997b). We address these suggestions based on the T1 sample.

The paper is organized as follows. 
In \S2.1 -- \S2.3 we summarize the creation of the T1 sample and the measurement of the AGN and host properties, analyzed in Papers I and II.  
In \S2.4 we describe the comparison type 2 sample we use, and account for differences in the measurement procedures. 
In \S3 we extend the relative decrease with $\lbol$ (the Baldwin effect) found in Paper II for \oiii\ and \Ha, to the \Hb, \nii, \sii\ and \oi\ lines. 
We then proceed in \S4 to measure the BPT ratios of the T1 sample, and their dependence on AGN and host characteristics. 
In \S5, we analyze objects which occupy a region in the BPT plots which is not populated in type 2 samples. 
In \S6, we identify the $M_* - Z$ relation in the T1 AGN. 
Analysis of LINERs and Composites is performed in \S7 and \S8. 
In \S9 we use the BPT ratios to constrain the AGN ionizing spectrum, and the location of the reddening dust. 
We summarize our results in \S10. 

Throughout the paper, we assume a FRW cosmology with $\Omega$ = 0.3, $\Lambda$ = 0.7 and $H_0 = 70\ \kms$ Mpc$^{-1}$.

\section{The Data}

\subsection{The T1 sample selection}
The T1 sample is selected from the 7\th\ data release of the Sloan Digital Sky Survey (SDSS DR7; Abazajian et al. 2009). The SDSS obtained imaging of a quarter of the sky in five bands ({\it ugriz}; Fukugita et al. 1996) to a 95\% $r$ band completeness limit of 22.2 mag. Objects are selected for spectroscopy mainly due to their non-stellar colors (Richards et al. 2002), or extended morphology (Strauss et al. 2002). The spectrographs cover the wavelength range 3800\AA --9200\AA\ at a resolution of $\sim150\ \kms$, and are flux-calibrated by matching the spectra of simultaneously observed standard stars to their PSF magnitude (Adelman-McCarthy et al. 2008).

We use SDSS spectra which have $0.005<z<0.31$ and are classified as quasars or galaxies. To ensure a reliable decomposition of the broad and narrow components of \Ha, we use only spectra with S/N $>10$ and a sufficient number of good spectral pixels in the vicinity of \Ha, as detailed in Paper I. These requirements are fulfilled by 232\,837 of the 1.6 million spectra in DR7, named here the parent sample. The spectra are corrected for foreground dust, using the maps of Schlegel et al. (1998) and the extinction law of Cardelli et al. (1989). Each spectrum is then fit with three galaxy eigenspectra representing the host (see \S2.2.4 below), and a $L_\lambda \propto \lambda^{-1.5}$ power law representing the AGN continuum. The host is subtracted, producing a spectrum free of stellar absorption features, excluding the Balmer absorption lines, which are handled at a later stage (see \S 2.2.4). We also subtract a featureless continuum, derived by 
interpolating the mean continuum level at 6125\AA--6250\AA\ and 6880\AA--7000\AA. The residual flux at 6250\AA--6880\AA\ ($\pm 14,000\ \kms$ from \Ha) is then summed, excluding regions $\pm 690\ \kms$ from the \oi\ $\lambda\lambda 6300,6363$, \nii\ $\lambda\lambda 6548,6583$, \sii\ $\lambda\lambda 6716,6731$ and \Ha\ narrow emission lines. We find 6\,986 objects with significant residual flux, which is potentially broad \Ha\ emission. 

For the objects with significant residual near \Ha, we fit the profiles of the broad and narrow \Ha, and the \oiii\ $\lambda 5007$, \oi, \nii\ and \sii\ doublets mentioned above. Narrow lines are fit using 4\th -order Gauss-Hermite functions (GHs; van der Marel \& Franx 1993) and an up to 10\th -order GH is used for the broad \Ha\ profile. Further details are given in \S2.4 of Paper I and \S\S2.3--2.4 of Paper II. 
The following criteria are applied to the broad \Ha\ fit, in order to exclude objects in which the residual flux is not clearly BLR emission: the FWHM ($\dv$) of the fit is required to be in the range $1\,000 - 25\,000\ \kms$; the total flux of the fit, and its flux density at the line centre, are required to be significant. As \oiii\ and \Hb\ are used extensively in this paper, we require them to have a sufficient number of good pixels in their vicinity for the fit to be reliable, as detailed in Paper I. 

Of the 3\,243 objects that pass these criteria, we use here 3\,175 objects in which our algorithm achieved reliable narrow line fits (see below). Due to the small relative number of objects in which the fitting algorithm did not succeed, we do not attempt to improve the algorithm further. The broad \Ha\ luminosity ($\lbha$) and $\dv$ of the 3\,175 objects of the T1 sample are listed in Table 1. The selection effects implied by our selection criteria are detailed in Paper I.

\begin{table}
\begin{tabular}{l|c|c|c|c|c|c}
Object name & $\lbha$ & $\dv$ & $M_*$ & $\luv$ & $\aox$ & Notes \\
\hline
J000202.95-103037.9  &  41.9  &  2310  &  10.9  &  43.8  &  -1.50  &  -,-,-  \\
J000410.80-104527.2  &  42.6  &  1360  &  11.0  &  44.6  &  -1.57  &  -,-,-  \\
J000611.55+145357.2  &  42.1  &  3320  &  11.1  &  44.0  &  -1.57  &  -,-,-  \\
J000614.36-010847.2  &  41.6  &  3910  &  10.7  &  43.2  &  -1.54  &  -,-,U  \\
J000657.76+152550.0  &  41.5  &  3020  &  10.0  &  43.0  &  -1.57  &  -,-,U  \\
\end{tabular}
\caption{The AGN and host characteristics of the T1 sample objects. The values of $\lbha$ and $\luv$ are in $\log\ \ergs$, $\dv$ is in $\kms$, and $M_*$ is in $\log\ \msun$. The last column lists notes for $M_*$, $\luv$ and $\lx$, separated by commas: `U' indicates an upper limit, and `N' indicates not available. The electronic version includes all 3\,175 T1 objects.}
\end{table}

\subsection{Narrow line measurements}

The narrow line luminosities of the T1 sample are listed in Table 2. We emphasize that these are luminosities within the SDSS 3\arcsec\ fibre, and that in all T1 objects the fibre was pointed at the centre of the host galaxy (see \S2.5 in Paper I). 
Below, we address the limitations of our fitting algorithm, which deblends the narrow lines from the broad lines and from the stellar absorption features. The success of the deblending can be further verified with higher S/N spectra, where the transitions between the different components are more prominent. Therefore, we corroborate our results by analyzing the mean spectra of different T1 subgroups, which have an effectively higher S/N.

\begin{table*}
\begin{tabular}{l|c|c|c|c|c|c|c|c|c|c|c|c}
 &  &  &  &  &  &  & &  & stellar & & & \\
Object name & \Hb & \oiii & \Ha & \nii & \sii & \oi &  Notes & robust & absorption  & BPT-\nii & BPT-\sii & BPT-\oi \\
\hline
J000202.95-103037.9  &  40.7  &  41.4  &  41.5  &  41.3  &  41.0  &  40.2  &  -,-,-,-,-,-  &  +  &  +  &        AGN  &  Seyfert  &  Seyfert  \\
J000410.80-104527.2  &  41.3  &  41.5  &  41.9  &  41.6  &    -1  &    -1  &  -,-,-,-,N,N  &  +  &  +  &  Composite  &       SF  &       SF  \\
J000611.55+145357.2  &  40.2  &  40.6  &  41.0  &  40.7  &  40.5  &  39.8  &  -,-,-,-,-,U  &  +  &  +  &  Composite  &       SF  &  Seyfert  \\
J000614.36-010847.2  &  40.4  &  40.9  &  41.1  &  40.7  &    -1  &  39.6  &  -,-,-,-,N,-  &  +  &  +  &  Composite  &       SF  &       SF  \\
J000657.76+152550.0  &  40.2  &  40.8  &  40.7  &  40.0  &  40.3  &  39.7  &  -,-,-,-,-,-  &  +  &  +  &  Composite  &  Seyfert  &  Seyfert  \\

\end{tabular}
\caption{The narrow line measurements of the T1 sample. 
All luminosities are in $\log\ \ergs$. 
Notes on the measurements of the six lines are separated by commas in column 8, ordered as in the table. A `U' designates an upper limit, an `N' designates bad pixels (for \sii\ and \oi\ only) which are also marked as $-1$ in the corresponding luminosity. 
Objects in which the \Ha, \Hb, or \nii\ narrow line measurements are not robust (\S 2.2.3), or the Balmer lines are affected by strong stellar absorption features (\S 2.2.4), are marked by an `x' in the respective following columns. Other objects are marked by a `+'. 
The last three columns list the classification of each object in the corresponding BPT panel. 
The electronic version includes all 3\,175 T1 objects.}
\end{table*}

\subsubsection{Bad pixels}

The main source of bad pixels in the SDSS spectra is poor sky subtraction, which degrades the spectrum mainly at $\lambda > 8000\AA$. Therefore, the \sii\ and \oi\ lines are not measurable in 612 (19\%) and 190 (6\%) of the T1 objects, respectively. These objects are marked in Table 2, and are disregarded in figures where the line is used. Objects in which one of the other lines used in this work has bad pixels do not enter the T1 sample (\S 2.1).

\subsubsection{Upper limits}

Our algorithm can robustly detect the six different narrow lines if their mean flux density $F_\lambda$ is 2--3.5 times the local flux density error. The exact value depends on how blended a specific line is with other spectral features, and is listed in Table 3. Upper limits on the fluxes of lines with lower $F_\lambda$ are derived by assuming a Gaussian profile, with a flux density equal to the minimum $F_\lambda$ required for detection and the width fit to the other narrow emission lines. Objects with upper limits are noted in Table 2. 

The T1 sample detection fractions of the different lines are listed in Table 3. The detection fractions are all $>77\%$.

\begin{table}
\begin{tabular}{c|c|c|c}
Narrow line & $\lambda$ (\AA) & min $F_\lambda/\epsilon_\lambda$ & Detection Fraction \\ 
\hline
\Hb   & 4961 & 3   & 0.84 \\
\oiii & 5007 & 3.5 & 0.99 \\
\oi   & 6300 & 2.5 & 0.77 \\
\Ha   & 6563 & 3   & 0.98 \\
\nii  & 6583 & 3   & 0.92 \\
\sii  & 6716 & 2.5 & 0.91 \\
\sii  & 6731 & 2   & 0.92 \\
\end{tabular}
\caption{The detection fractions of the narrow lines used in the paper. Col. 3 notes the minimum flux density required for detection, in units of the local flux density error.}
\end{table}

\subsubsection{\oiii-like narrow lines}

As noted in Papers I and II, in 15\% of the sample the fit yielded FWHM(n\Ha) $\geq 1.5 \times$ FWHM(\oiii). These objects have non- or barely-detectable narrow lines near \Ha, and there is no clear transition between the broad and narrow components of the Balmer lines. Therefore, we fit the narrow lines near \Ha\ in these objects with a FWHM, 3\rd\ and 4\th\ GH parameters equal to those found for \oiii. 

An eye-inspection of the narrow \Hb\ fits yielded another 188 objects (6\%) without a clear NLR/BLR transition, despite having FWHM $<1.5\times$ FWHM(\oiii). We refit these objects with \oiii-like profiles, and updated the relevant narrow line fluxes. The new fit failed in 68 of the objects (reduced $\chi^2>2$). Due to their relatively small number, we did not attempt to improve the fit, and simply removed these 68 objects from the sample. 
This change in the narrow line fluxes of 6\% of the T1 sample has a negligible effect on the results presented in Papers I and II. 

The narrow \Ha, \Hb, and \nii\ line fluxes are less certain in objects fit with an \oiii-like profile. Therefore, throughout the paper different symbols are used when these measurements are utilized. These objects are also noted in Table 2.

\subsubsection{Strong stellar Balmer absorption}

We model the stellar absorption features by fitting the first three Yip et al. (2004) eigenspectra (ESa) to the SDSS spectra, together with a power law for the AGN continuum. Since the Yip et al. ESa have emission lines, in ES1 we replace the lines with the absorption features of the Hao et al. (2005) ES1 (detailed in \S2.2 of Paper II). This step is justified since both ES1's represent an old stellar population. In ES2 and ES3, which represent a younger population, an emission line free ES is not available, so we simply interpolate over the lines. 

Since the absorption lines are significantly wider than the emission lines, an interpolation over the emission will not remove the entire absorption feature. However, near \Ha\ the interpolation is done also over the \nii\ lines which flank \Ha. Therefore, our fit does not account for the entire \Ha\ absorption feature of young stars. In Paper II, we found that in the 5\% of the T1 objects that have $\lnha < 3\AA \times L_\lambda$(host), the $\lnha$ are underestimated due to improper subtraction of the stellar absorption. 
Now, the narrow \Hb\ emission line is weaker than \Ha, and therefore more suspect to significant biases due to improper subtraction of the stellar absorption features. However, near \Hb\ the interpolation in ES1 and ES2 is performed only over the narrow \Hb\ line, so the wide part of the stellar absorption feature is accounted for by our fit. 
Therefore we mark the same objects as in Paper II, i.e. objects with $\lnha < 3\AA \times L_\lambda$(host), as objects with potentially underestimated $\lnha$ and $\lnhb$. 
We verify below this suffices in order to identify objects with offset $\lnhb$ values. 

\subsection{Additional Properties}

\subsubsection{$\lhost$ and $M_*$}

We derive the host galaxy luminosity, $\lhost$, by subtracting the estimated net AGN luminosity from the total observed luminosity. 
For the total observed luminosity we use the SDSS {\sc cModel} flux\footnote{Not available for seven objects. They are disregarded when $M_*$ is used.} (Abazajian et al. 2004) in the $z$-band, which is a linear sum of a de Vaucouleurs model and an exponential model fit to the image, and is the best suited model to account for both the galaxy and the nuclear light. 
The $z$-band is chosen since it is the reddest SDSS band, therefore it has the highest host to AGN contrast. It also has the smallest dispersion in the ratio of host mass to host light. We estimate the net AGN luminosity at the $z$-band, $L_{{\rm AGN};\ z-{\rm band}}$, to be $10 \cdot \lbha$ (Paper I). We do not use the eigenspectra fit described in \S2.1 to estimate the host luminosity, due to degeneracies between the host and AGN continuum flux in this fit (see \S 2.2 in Paper I).

To convert the $\lhost$ of the T1 AGN to $M_*$, we compare $M_*$ with $L_{z-{\rm band}}$ in the type 2 AGN sample described below. The $M_*$ of the type 2 AGN were measured by Kauffmann et al. (2003b), as part of the MPA/JHU analysis of SDSS spectra\footnote{Available at http://www.mpa-garching.mpg.de/SDSS/DR7/}. Also, Ka03 found that the mean color of type 2 AGN hosts becomes bluer with increasing $\loiii$. Accordingly, we calculate the mean mass to $z-{\rm band}$ light ratio for each $\loiii$ (in 0.5 dex bins), and find a mean $M/L=2.6$ at $\loiii=10^{39}\ \ergs$ and $M/L=1.7$ at $\loiii=10^{42.5}$, where $M/L$ is given in solar units. The $M/L$ dispersion in each $\loiii$ bin is $\sim 0.15$ dex. 
In Paper I, we showed that the color of the mean hosts of type 1 AGN at different luminosities equals the mean color of type 2 hosts with the same luminosity. 
Therefore, for each T1 AGN we use the $M/L$ appropriate for its $\loiii$. We note that if we had used the median $M/L$ for all T1 AGN, the implied $M_*$ would have changed by $<0.1$ dex. 
The individual $M_*$ of the T1 sample objects are listed in Table 1.

An additional source of error is the scatter in the ratio of $L_{{\rm AGN};\ z-{\rm band}}$ to $\lbha$. 
We assume this scatter equals the scatter in the relation between $L_{{\rm total};\ 5100\AA}$ and $\lbhb$ of $0.5<z<0.7$ SDSS quasars -- the lower $z$ limit ensures the quasars are luminous and host contribution to the continuum is minimal, while the upper $z$ limit ensures \Hb\ fully appears in the spectrum. Using the $\lbhb$ and $L_{{\rm total;} \ 5100\AA}$ values from Shen et al. (2011), we find a scatter of 0.2 dex. 
This scatter implies that in the 9\% of the T1 objects with implied $\lagn/\lhost > 1$, the true $\lhost$ may be overestimated by a factor of more than 2, therefore we treat these measurements of $\lhost$ as upper limits. 
In the 3\% of T1s with implied $\lagn/\lhost>3$, the true $\lhost$ may also be underestimated by a factor of more than 2. In 0.5\% of the objects, the implied $\lhost$ is negative. In both cases we set $\lagn/\lhost = 3$, and treat these measurements of $\lhost$ as upper limits.

\subsubsection{$\luv$ and $\aox$}

We derive the $\luv\ (\equiv \nln(1528\AA))$ and $\lx\ (\equiv \nln(2~\kev))$ of the T1 AGN, from the GALEX (Martin et al. 2005) and ROSAT (Voges et al. 1999) surveys. GALEX observed 89\% of the T1s, and detected 93\% of them. ROSAT observed the entire sky, and detected 43\% of the T1s. The derivation of the luminosities is detailed in Paper II. Table 1 lists $\luv$ and $\aox \equiv -0.42\times\log \luv/\lx$, the slope of the interpolated power law between the UV and the X-ray.

\subsection{The T2 sample}

We compare our results to the Brinchmann et al. (2004) type 2 AGN sample, which was derived from the SDSS galaxy survey, using the emission lines measurement of the MPA/JHU group. The type 2 AGN were selected by requiring S/N $>3$ in the \oiii, \Hb, \nii, and \Ha\ narrow emission lines, and being above the Ke01 `extreme starburst' line in the BPT-\nii\ panel. We use all type 2 objects that appear in our parent sample (following the S/N $>10$ and bad pixel cuts, \S 2.1), excluding the 454 objects which enter the T1 sample, as they show broad \Ha\ emission. We name these 13\,705 objects as the T2 sample.

The MPA/JHU group modeled the stellar absorption features using the Bruzual and Charlot (2003) stellar library. We use a simpler technique in the T1 sample, based on the Yip et al. (2004) ESa, due to possible degeneracies of different stellar components with the unobscured AGN continuum (see \S2.2 in Paper I). In order to understand the effect of the different stellar modeling techniques on the measured narrow line ratios, and the effect of other differences in the fitting procedure, we run our fitting algorithm on 700 spectra from the T2 sample\footnote{The first 700 objects, sorted by right ascension.}. Then, we compare the narrow line ratios we measure on these T2s with those published by the MPA/JHU group.

In these 700 type 2s, our algorithm gives \oiii/\Hb\ ratios which are on average 0.08 dex larger than the ratios measured by MPA/JHU, with a dispersion of 0.11 dex. Our \nii/\Ha, \sii/\Ha\ and \oi/\Ha\ measurements are on average 0.04, 0.1 and 0.07 dex larger than MPA/JHU, with dispersions of 0.09, 0.1, and 0.1 dex. 
The offsets in the narrow line ratios are mainly due to offsets in the measured flux of the narrow \Ha\ and \Hb\ lines (mean offset $-0.09$ dex each), which could imply that we did not fully correct for the stellar Balmer absorption features. 
Therefore, to minimize offsets between the T1 and T2 sample which originate from measurement issues, we hitherto decrease the BPT ratios we measure in the T1 sample objects by these mean offsets. 
Also, we assess the systematic error in our measurement of these ratios to be 0.1 dex.

\section{The Baldwin effect of the narrow lines}

In Figure 1, we present the ratio of the narrow lines luminosity $\lnhb$, $\lnii$, $\lsii$, and $\loi$, with $\lbha$  as a function of $\lbha$. 
Black dots mark objects with robust measurements, while gray markers indicate the less robust values. 
For each narrow line, we perform a least-squares best fit of $L_{\rm NL}$ vs. $\lbha$, where $L_{\rm NL}$ is the luminosity of the narrow line. 
We treat $\lbha$, which is used to select the T1 sample, as the independent variable. 
We find $\lnhb \propto \lbha^{0.67}, \lnii \propto \lbha^{0.54}, \lsii \propto \lbha^{0.53}$, and $\loi \propto \lbha^{0.63}$, with dispersions in the range $\sigma=0.32-0.38$. The formal error on all slopes is $\sim0.01$. 
A significant trend of decreasing NLR to BLR luminosity ratio with increasing  $\lbha$ is clearly seen for all lines. 

In Paper I, we found that the observed mean optical-UV SED of the T1 sample is well matched by a fixed shape SED of luminous quasars, which scales linearly with $\lbha$, and a host galaxy contribution. Therefore, $\lbha \propto \lcont$, where $\lcont$ is the AGN continuum luminosity near \Ha, and the trends observed in Figure 1 represent a Baldwin effect (Baldwin 1977) for the narrow lines. 

However, we note that even if intrinsically $L_{\rm NL} \propto \lcont ^ {1.0}$, i.e. no intrinsic Baldwin effect, then due to the dispersion in $\ewha\ (\equiv \lbha / \lcont)$ we expect to find $L_{\rm NL} \propto \lbha^{1-\epsilon}$. In Appendix C, we show that $\epsilon \leq (\frac{\sigma(\ewha)}{\Delta(\lbha)})^2$, where $\sigma(\ewha)$ is the intrinsic dispersion in $\ewha$, and $\Delta(\lbha)$ is the standard deviation of the distribution of $\lbha$ spanned by the sample. 
In the T1 sample we have $\Delta(\lbha)=0.75$, and we assume that $\sigma(\ewha) = 0.2$ dex, as found for quasars (\S2.3.1). Therefore, $\epsilon \leq (0.2/0.75)^2 = 0.07$. This $\epsilon$ is significantly smaller than the slopes of $\gtrsim 0.3$ found above, indicating that the observed trends in $L_{\rm NL}/\lbha$ indeed 
represent intrinsic Baldwin effects.

The relations found in Paper II for \oiii\ and \Ha\ are $\lnha \propto\lbha^{0.67}, \sigma=0.37$ and $\loiii\propto\lbha^{0.72}, \sigma=0.36$.
Note that the different slopes found above imply some trends in the mean positions with luminosity of the T1 objects in the BPT plots, as shown below.

\begin{figure}
\includegraphics{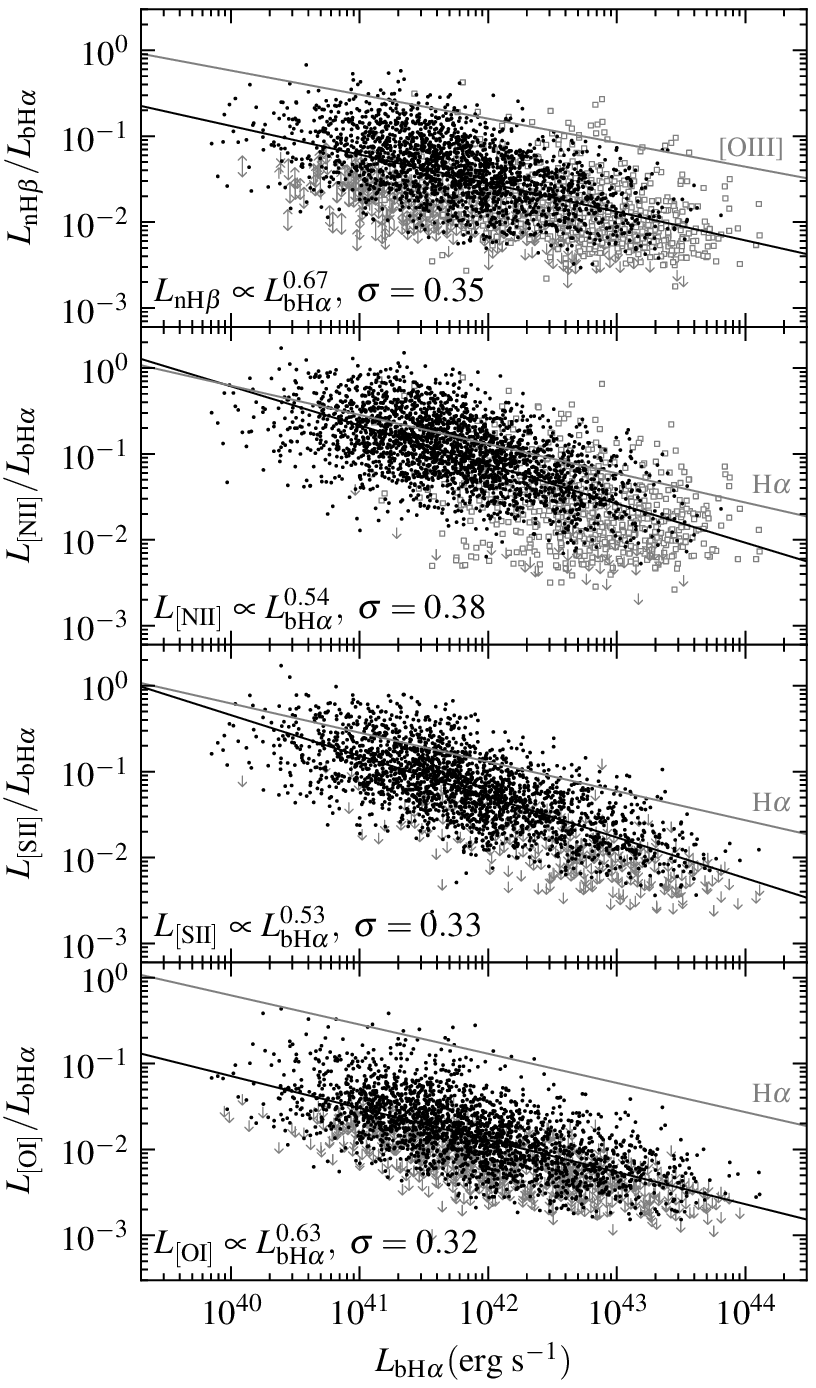}
\caption{The distribution of $L_{\rm narrow\ line}/\lbha$ versus $\lbha$ for narrow lines
analyzed in the BPT plots below. 
Robust narrow line measurements are marked by black dots. 
Profiles of \nii\ and \Hb\ based on \oiii\ (\S 2.2.3) are marked by gray squares. 
Upper limits (non detections, \S2.2.2), and lower limits (stellar absorption for \Hb\ only, \S2.2.4) are marked by the appropriate arrows. 
The slope of the best fitting power laws (black lines) and the associated dispersion are noted. For comparison, the relations found in Paper II, for \oiii\ (index = 0.72) and for \Ha\ (index = 0.66), are shown as gray lines. 
The tendency of increasing NLR / BLR ratio with decreasing $\lbha$, found in Paper II, is seen in all lines. Since $\lbha \propto \lbol$ (Paper I), the observed trends represent the Baldwin effect of the narrow lines. 
When excluding objects marked by squares, the \nii\ Baldwin slope increases to 0.60. 
Note that \sii\ and \nii\ have steeper slopes than \Ha, which imply a shift in the mean positions in the BPT plots with luminosity.
}
\label{fig: }
\end{figure}

\subsection{Less robust values}

In all four panels of Figure 1, most upper limits fall within the distribution of the general population. 
The objects in which the narrow lines are fit with an \oiii-like profile (\S2.2.3) are located at the high-$\lbha$ end of the sample. As noted in \S2.2.3, the deblending of the Balmer lines and \nii\ from the broad lines may be inaccurate in these objects. Indeed, the $\lnii/\lbha$ values of these objects are offset to lower values then the general trend. When excluding these objects, we find $\lnhb \propto \lbha^{0.68}$ and $\lnii \propto \lbha^{0.60}$, i.e. a similar \Hb\ slope and a \nii\ slope higher by 0.06 compared to when using all objects. 

\subsection{Comparison with previous studies}

Croom et al. (2002) compared the narrow \oiii, \oii, \neiii, and \nev\ line luminosities with the absolute B magnitudes of 2dF and 6dF quasars (Croom et al. 2001). 
For a direct comparison with our results we subtract the slope they found for each line with the positive slope
of $+0.18$ they found for $\lbhb$. Comparing the narrow lines to the broad \Hb\ also avoids the bias created by host contamination of the continuum. This contamination likely creates the inverse Baldwin relation 
(i.e. positive slope) for the broad \Hb\ line found
by Croom et al., in contrast with the absence of a Baldwin relation (i.e. zero slope) for the Balmer lines 
found in our earlier analysis (Paper I). 
Their implied narrow lines versus broad \Hb\ slopes are 0.86, 0.49, 0.58 and 0.74 for \oiii, \oii, \nev, and \neiii, respectively. All their narrow lines show a Baldwin effect, as found here. 
Their \oiii\ slope of 0.86 is steeper then our slope of 0.72, while their \oii\ slope of 0.49 is flatter than 
our flattest slope of 0.53 for \sii. 

Very recently, Zhang et al. (2012) compared narrow line equivalent widths with the continuum luminosity at 5100\AA\ in mean spectra of SDSS type 1 AGN. As with Croom et al. above, we subtract the slope of $+0.16$ (see \S3.1 in Zhang et al.) they found for $\lbhb$ from the slope they found for each line. The implied slopes are $-0.45,\ -0.44,\ -0.26,\ -0.36,\ -0.32$ and $-0.37$ for the narrow \Ha, \Hb, \nii, \sii, \oi\ and \oiii, respectively. 
The implied Zheng et al. Baldwin slopes of all lines except \nii\ differ by $\lesssim 0.1$ from the slopes found here. The higher value of $0.2$ in the slope of \nii\ could be because \nii\ increases with $M_*$, and $\lbol$ and $M_*$ are correlated in the Zhang et al. sample, but not in the T1 sample (see below).

H{\"o}nig et al. (2008) and Keremedjiev et al. (2009) showed that mid-IR narrow lines also show Baldwin effects.

\section{The BPT positions of the T1 AGN}

Figure 2 presents the BPT positions of the 3\,175 T1 AGN, plotted over the SDSS narrow line galaxies (Figure 1 from Ke06). Classification lines are from Ke01, Ka03, Ke06 and Ho et al. (1997a, hereafter Ho97). The classification of each T1 object in each panel is listed in Table 2.

\begin{figure*}
\includegraphics{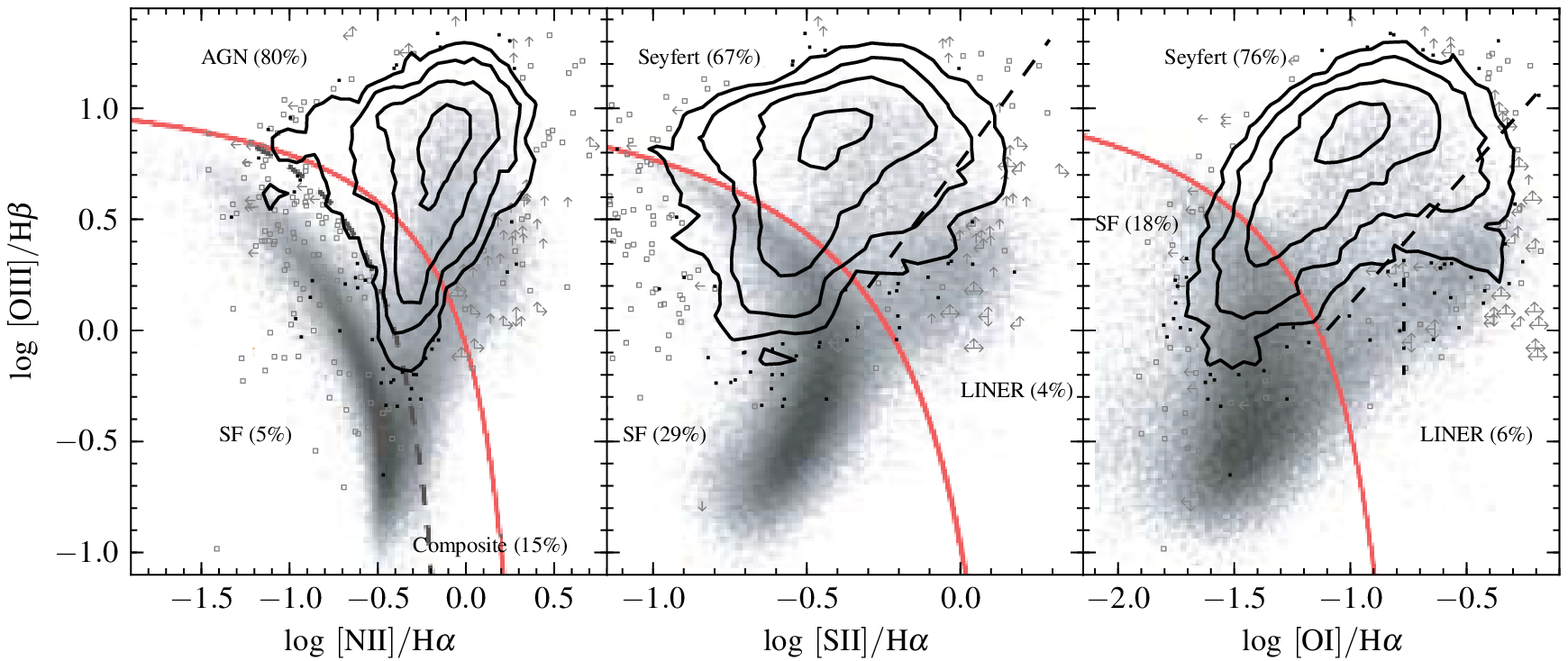}  
\caption{The BPT positions of the 3\,175 T1 AGN (solid contours), compared to the narrow line SDSS galaxies (background gray pixels, Figure 1 in Ke06). Red solid lines are the Ke01 extreme starburst lines, while the dashed lines mark the separation between `SF' and `Composites' in the \nii\ panel (Ka03), and between `Seyferts' and `LINERs' in the \sii\ and \oi\ panels (Ke06). The dash-dotted line in the \oi\ panel further divides the LINER group into bona-fide LINERs and `transition' objects (Ho97). 
The contours encircle regions with 10, 25, 60 and 150 T1 objects per 0.2x0.2 dex$^2$ bin, respectively.
Beyond the outer contour, T1 AGN are marked as in Figure 1, with upper / lower limits on either of the emission lines denoted by an appropriate arrow. 
In the \nii\ panel, 20\% of the T1 objects are below the Ke01 line, and would not be defined as AGN. Of these, 15\% are defined as Composites, and 5\% as SF galaxies. 
In the \oi\ panel, 190 objects in the sample are LINER 1s, of which 94\% (179/190) appear to the right of the Ho97 line. 
Note that the T1 extends to higher \oiii/\Hb, and lower \nii/\Ha\ and \sii/\Ha,  
compared to the narrow line galaxies and AGN.}
\label{fig: }
\end{figure*}

The T2 AGN reside, by definition, above the Ke01 line in the BPT-\nii\ panel. 
However, only 80\% of the T1 objects reside in the AGN regime, 15\% are classified as composite 
and 5\% as SF. We stress again that all T1 AGN are clearly powered by accretion onto a massive black 
hole, as indicated by the detection of a broad \Ha. Thus, the SDSS type 2 AGN sample is likely
only 80\% complete. Including composites will increases the completeness to 95\%, but may
include a significant number of objects which are not true AGN. 

We note that the narrow line measurements of two-thirds of the T1s which reside in the SF region are poorly constrained. Thus, with higher quality spectra the true AGN fraction with SF narrow line ratios may therefore be as low as 2\%. In comparison, only 18\% of the T1s classified as composites and 17\% of the T1s classified as `AGN' have poorly-constrained narrow line measurements. 

The fraction of T1 which reside outside the AGN region in the BPT-\oi\ panel is 18\%, and in the
BPT-\sii\ panel it reaches 29\%. The SDSS spectra are taken with a 3'' fiber, which can include a significant fraction of the host galaxy emission. Below we study some indications that the offset
from the AGN region in the BPT plots indeed results from host contamination.

Figure 2 also shows that a sizable fraction of the T1 sample occupies a new region in the BPT panels, with \oiii/\Hb $= 5-10$, \nii/\Ha $=0.1-0.3$, and \sii/\Ha\ = $0.1-0.3$. These objects have no counterpart in the narrow line sample. Specifically, 10\% of T1s with $\oiii/\Hb>5$ have $\nii/\Ha<0.3$, compared to only 0.8\% of the T2 sample. This result is consistent with the Zhang et al. (2008) result. 
Below, we study the range of AGN and host properties at which these line ratios are dominant, and discuss their physical origin.

In the \oi\ panel, 190 objects in the T1 sample are classified as LINER 1s. The vast majority (179) of them appear to the right of the Ho97 line, where the `bona-fide' LINERs reside. This result is consistent with the strong drop in broad \Ha\ detection across this line (Ho et al. 1997b, Ho 2008, Wang et al. 2009).

\subsection{BPT positions of T1 AGN, by $\lbha$, $\lledd$ and $\lagn/\lhost$}

In this section we utilize the large size of the T1 sample, and explore their positions within the
BPT plots when the sample is cut based on various AGN and host properties. We identify some qualitative trends, which are further explored in the following sections. 

Figure 3 presents the BPT positions of the individual T1 objects as a function of $\lbha$, which is a measure of $\lbol$ ($=130\times \lbha$, Paper I). As in Figure 2, the positions of the T1s are plotted over the SDSS narrow line galaxies from Ke06.
At $\log \lbha = 40.7\ (\lbol = 42.8)$, T1 AGN largely overlap the narrow line sample. With increasing luminosity, the T1 AGN shift to lower \nii/\Ha, lower \sii/\Ha\ values, slightly lower \oi/\Ha, and
higher \oiii/\Hb\ values, as expected from the different luminosity trends of the different lines (Figure 1).
At quasar luminosities ($\log\ \lbha \gtrsim 43$ or $\log \lbol \gtrsim 45$), the T1 distribution is distinct from the type 2 distribution in the \sii-panel, and is offset in the \nii-panel to lower values.
Also, the fraction of AGN which reside below the Ke01 line decreases with increasing $\lbha$.

\begin{figure*}
\includegraphics{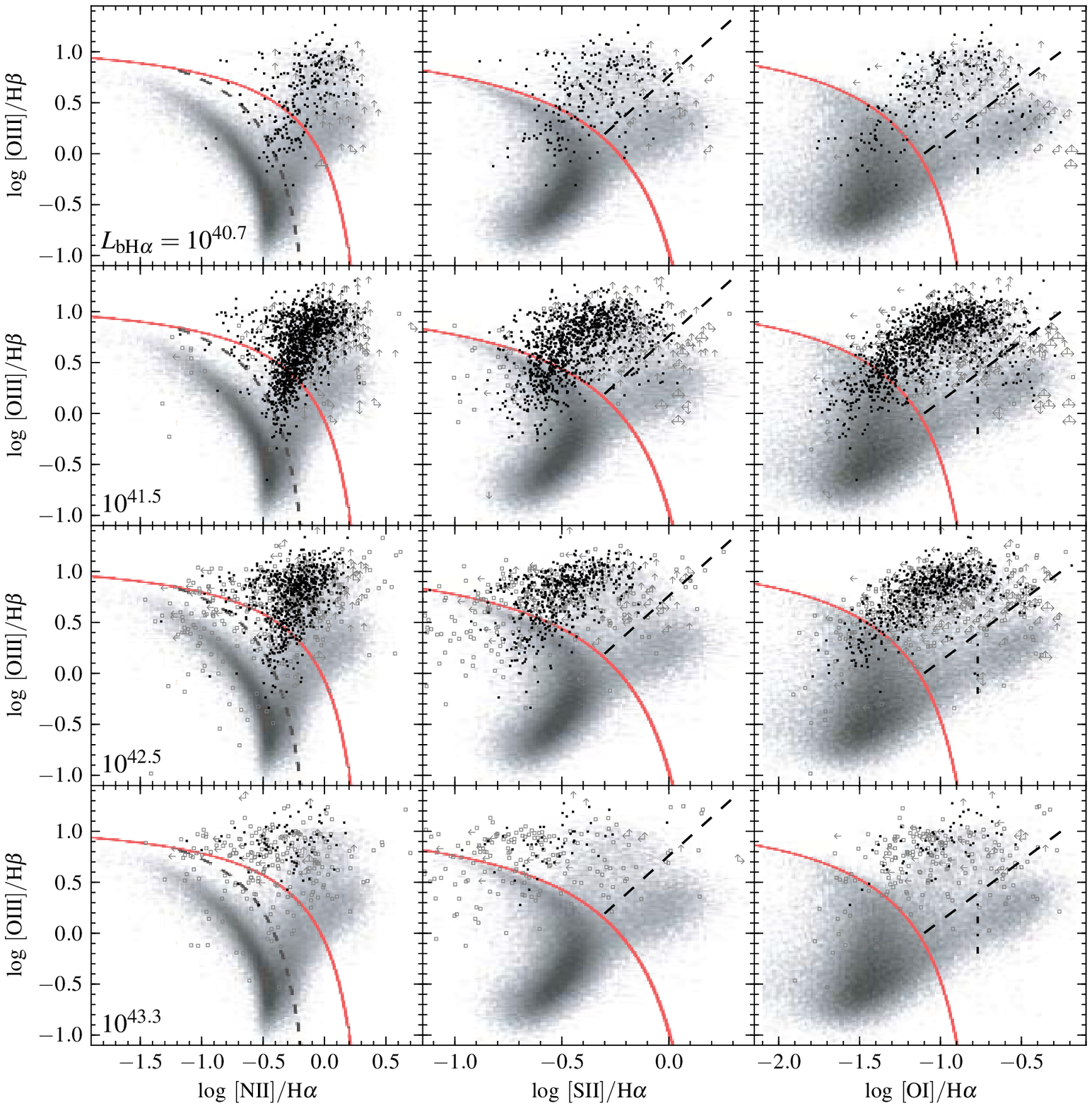}
\caption{The dependence of the BPT position on the AGN luminosity. The T1 sample markers are as in Figure 1. The narrow line background and dividing lines are described in Figure 2. Each row presents T1 AGN from a given decade-wide bin in $\lbha$. The mean $\lbha$ in each bin are noted (in $\ergs$). At $\log \lbha = 40.7\ (\lbol = 42.8)$, T1 AGN overlap the narrow line sample.  
With increasing luminosity, the mean $\nii/\Ha$ and $\sii/\Ha$ decrease. At quasar luminosities ($\log\ \lbha \gtrsim 43$ or $\log \lbol \gtrsim 45$), a large fraction of the T1 AGN occupy a region in the \nii\ and \sii\ panels which is distinct from the type 2 distribution.
}
\label{fig: }
\end{figure*}

The fraction of poorly-constrained objects increases with $\lbha$ (Figure 1), due to the decrease in the
relative strengths of the narrow lines (Paper II and Figure 1). Therefore, one may wonder whether this trend with luminosity is not simply due to the limitations of the deblending algorithm. In appendix A we verify the observed trend using high quality mean spectra. 

Figure 4 shows the BPT positions of the T1 objects, now subdivided by $\lledd$ (derived from $\lbha$ and $\dv$, using eq. 3 in Paper I). 
Several trends in the BPT position are apparent. With increasing $\lledd$, an increasing fraction of T1s have high \oiii/\Hb, low \nii/\Ha, and low \sii/\Ha, as found with increasing $\lbha$ in Figure 3. The decrease in \oi/\Ha\ with $\lledd$ is more pronounced than in Figure 3: the \oi/\Ha\ span mainly 0.1 -- 0.3 at low $\lledd$, compared to 0.03 -- 0.1 at high $\lledd$. The fraction of LINERs in the BPT-\oi\ panel strongly decreases with increasing $\lledd$, from 29\% at $\log\ \lledd = -2.5$, to 6\% at $\log\ \lledd =-1.8$ and 3\% at $\log\ \lledd = -1.2$ and $-0.6$. 

\begin{figure*}
\includegraphics{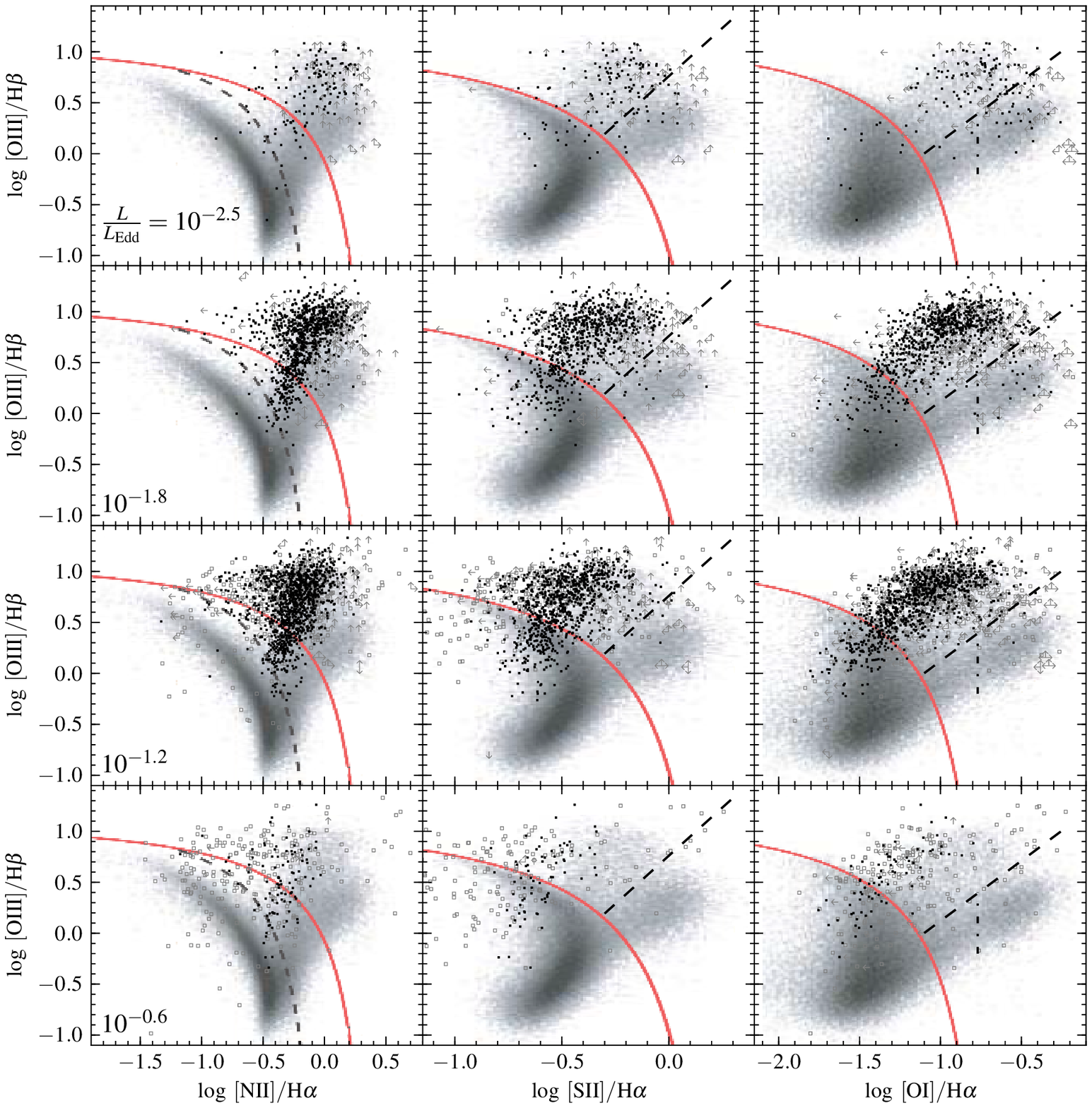}
\caption{As Figure 3, for the dependence of the BPT position on $\lledd$. 
Each row presents T1 AGN for a given $\lledd$ bin (0.75 dex wide). The mean $\lledd$ in each bin is noted in the lower left corners. With increasing $\lledd$, the T1 AGN move to the left in all BPT panels. Also, at $\log\ \lledd = -2.5$, 29\% of the T1 sample are in the LINER region of the \oi\ panel, compared to 6\% in the entire T1 sample. 
}
\label{fig: }
\end{figure*}

Figure 5 is similar to Figs. 3 and 4, with different rows designating different bins in $\lagn/\lhost$, measured at the SDSS-$z$ band (see \S 2.3.1). 
We note the division of objects between the two high $\lagn/\lhost$ bins is not robust in objects with $\lagn/\lhost>1$, due to the possible error in the determination of $\lhost$. The 24 objects with a negative implied $\lhost$ appear in the $\lagn/\lhost\geq 2$ bin. 
With increasing $\lagn/\lhost$, \nii/\Ha\ and \sii/\Ha\ decrease, as found with increasing $\lbha$ in Figure 3, and with increasing $\lledd$ in Figure 4. The T1 sample overlaps the type 2 sample in host dominated objects, and is distinct from the type 2 distribution in AGN dominated objects. Also, the composite fraction decreases from 22\% at $\lagn/\lhost=0.04$ to 6\% at $\lagn/\lhost\geq 2$. 

\begin{figure*}
\includegraphics{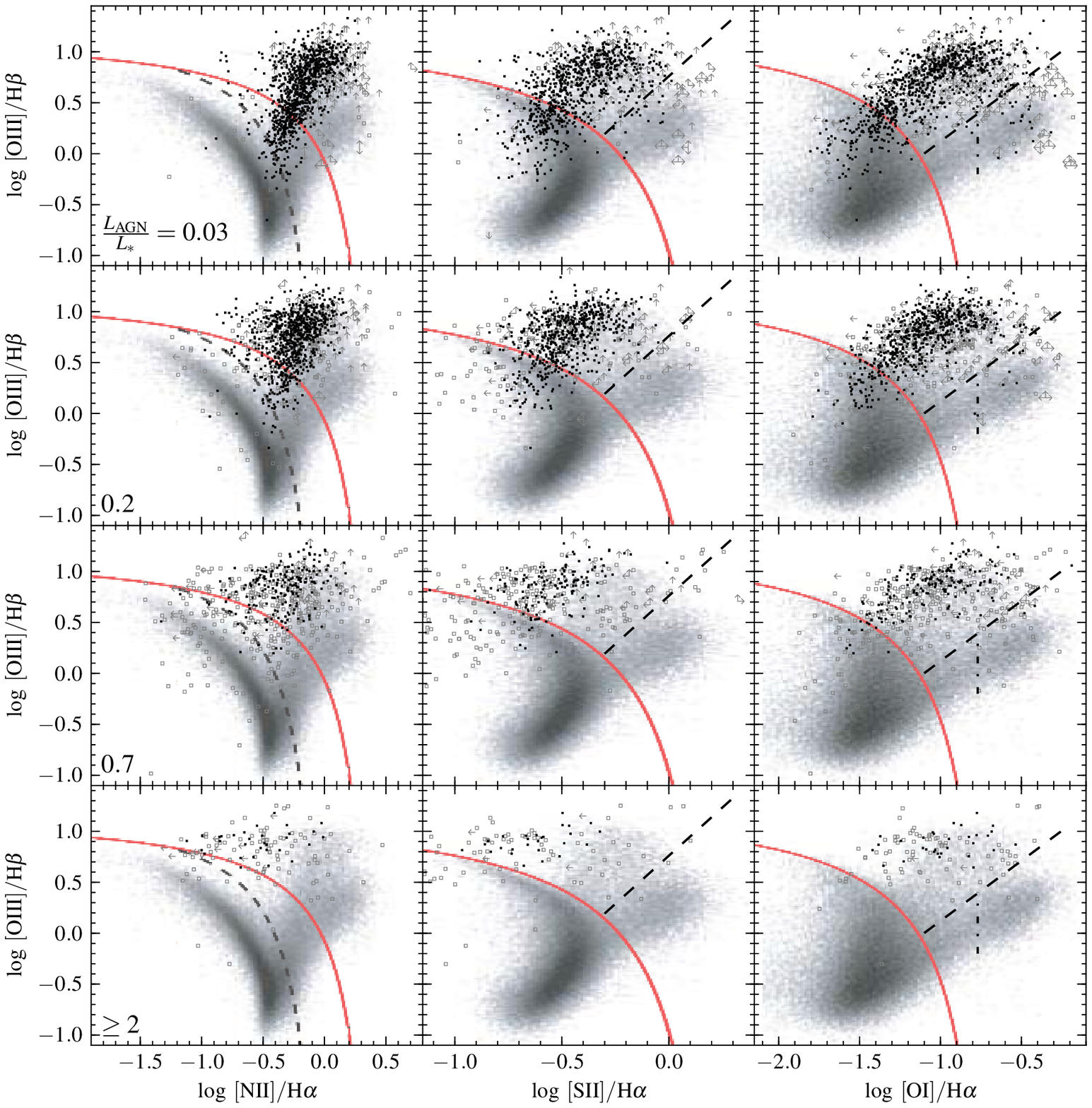}
\caption{
As Figure 3, for the dependence of the BPT position on $\lagn/\lhost$.
Each row presents T1 AGN from a given  $\lagn/\lhost$ bin (0.7 dex wide). The $\lagn/\lhost$ is calculated at the SDSS $z$-band, and the mean values are noted in the lower left corners. 
The T1 AGN move to lower \nii/\Ha\ and \sii/\Ha\ with increasing $\lagn/\lhost$, similar to the trend with $\lbha$ seen in Figure 3. 
The composite fraction decreases from 22\% at $\lagn/\lhost=0.03$ to 6\% at $\lagn/\lhost\geq2$.
}
\label{fig: }
\end{figure*}

To satisfy the curious reader, the mass dependencies are explored in Appendix B, where we plot the BPT positions of the T1 objects, subdivided by $\mbh$ and $M_*$.

\subsection{Comparison with different type 1 samples}

Here, we compare the BPT positions of the T1 sample and its dependence on AGN and host properties (Figs. 2 -- 5), with NLR studies of other type 1 AGN samples, which were selected differently.

Greene \& Ho (2007) inspected the narrow line ratio of 229 SDSS type 1 AGN, selected based on the detection of a broad \Ha, similar to T1, but required to have $\mbh < 2\times 10^6\ \msun$. In their sample, 39\% of the objects are classified as Composites or SFs in the BPT-\nii\ panel, versus
only 20\% (Figure 2) in our sample. However, when we restrict the T1 sample to 
$\mbh < 2\times 10^6\ \msun$ (Figure B1, upper panel), the fraction increases to 36\%, consistent
with the Greene \& Ho (2007) result. In a followup paper (Xiao et al. 2012), 
they compared the narrow line ratios based on the SDSS spectra with ratios based on spectra from a smaller aperture. The fraction of Composites / SFs decreased to 18\%, indicating that extended emission from SF in the host galaxy shifts the BPT position into the composite region at low $\mbh$. Below, we provide further evidence that this effect applies also to composites at higher $\mbh$.

Another prominent feature in the Greene \& Ho low $\mbh$ sample are objects with high \oiii/\Hb, low \sii/\Ha\ and low \nii/\Ha. Of their type 1's with $\oiii/\Hb>5$, 10\% have $\nii/\Ha<0.3$, as found for the entire T1 sample above. These narrow line ratios are also observed in low $\mbh$ type 2 samples (Barth et al. 2008), but are clearly missing from type 2s at higher luminosity (Figure 2). It therefore seems that type 1 and type 2s have similar ratios at low $\mbh$, but become distinct at higher $\mbh$. We examine the reason for this difference below.

Winter et al. (2010) published the BPT positions of a hard X-ray selected AGN sample, of which they identified 33 objects as broad line AGN\footnote{They excluded Sy1.8s and Sy1.9s, which selects against low luminosity type 1 AGN (Paper II).}. Their mean log $\loiii$ is a factor of three higher than the mean in the T1 sample. 
Five of their type 1s have \sii/\Ha\ $<0.1$, and two have \nii/\Ha\ $<0.1$, which are not seen in their type 2 sample. These ratios are seen in the T1 sample, but not in the SDSS narrow line sample.

Buttiglione et al. (2010) measured the BPT positions of a radio selected AGN sample. They show a clear trend of decreasing \nii/\Ha\ and \sii/\Ha, and increasing $\oiii/\Hb$, with increasing \oiii\ luminosity (figure 1 there). Their trend is equivalent to the trend seen in Figure 3 with $\lbha$. A similar trend can be seen in figure 4 of Wang \& Wei (2010), who measured the BPT ratios of Seyferts 1.8s and 1.9s with ROSAT detections. They found that objects in which the AGN dominates the continuum are offset to lower \nii/\Ha\ than objects in which the continuum is host dominated, as can be seen here in Figure 5.

\begin{figure}
\includegraphics{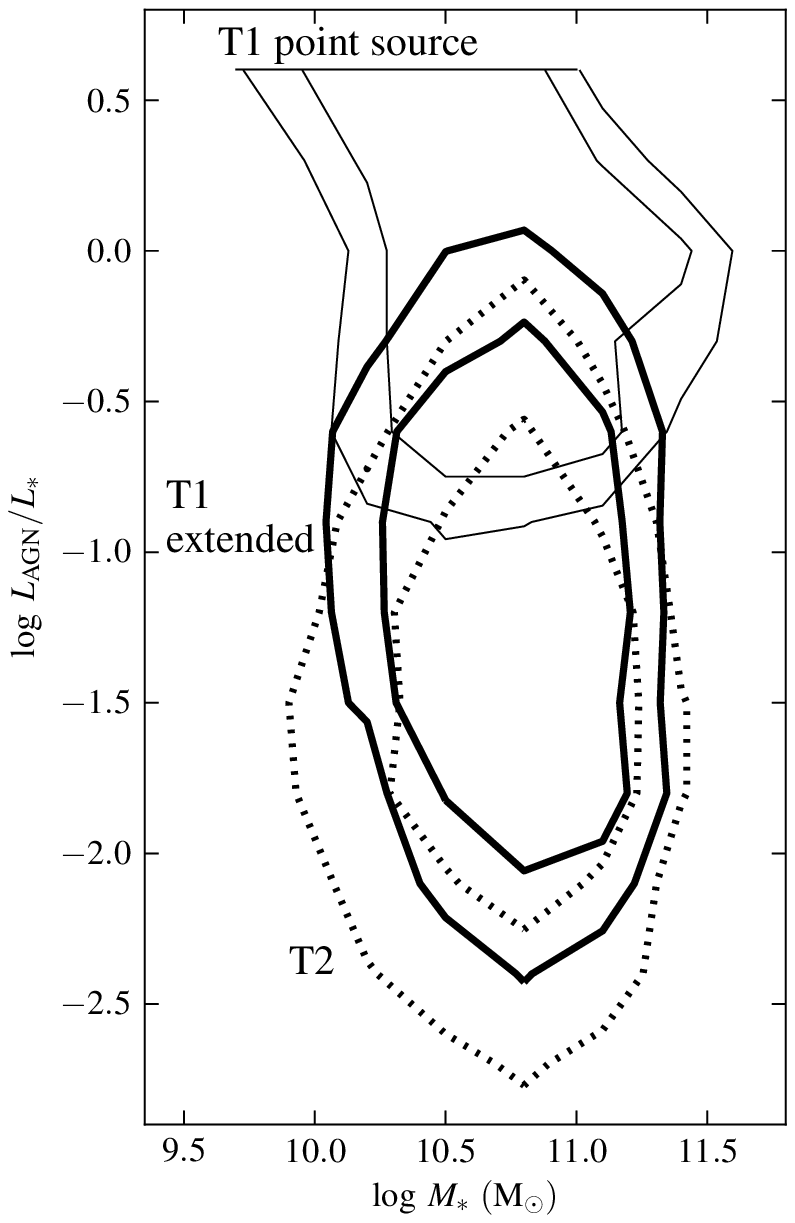} 
\caption{The distribution of the T1 and T2 AGN, with $\oiii/\Hb>5$,
in the $M_*$ versus $\lagn/\lhost$ plane. The distributions are denoted by two contour lines, which indicate the number of objects per $0.3\times 0.3\ {\rm dec}^2$, 1\% of the size of the group
for the outer contour, and 2.5\% for the inner contour. The three groups marked are T1 selected as point sources (thin solid), T1 selected to have extended morphology (thick solid), and T2s which are all selected to have an extended morphology (dotted). 
The maximum $\lagn/\lhost$ found by our algorithm is 3. At $\lagn/\lhost>1$, $\lhost$ (and $M_*$) may be overestimated (\S2.3.1). 
The extended T1 distribution overlaps the T2 distribution, as expected from their common selection criteria (and AGN unification). 
The T1 point sources, selected by their non-stellar colors, are offset to higher $\lagn/\lhost$ than AGN with an extended morphology, and constitute an AGN population which does not appear in the T2 sample. 
}
\label{fig: }
\end{figure}

\section{The offset of T1-AGN to low \nii/\Ha\ and \sii/\Ha}

Figs. 2--5 show that the distributions of the narrow line ratios of the T1 objects extend to values
which are not seen in the SDSS narrow line sample, in particular at high $\lbol$, high $\lledd$, and high $\lagn/\lhost$. These non overlapping objects have \oiii/\Hb\ similar to type 2s, but lower \nii/\Ha\ and \sii/\Ha. In Appendix A we show the offset ratios are not an NLR/BLR deblending artifact. Why are these ratios absent from type 2 samples? 

\begin{figure}
\includegraphics{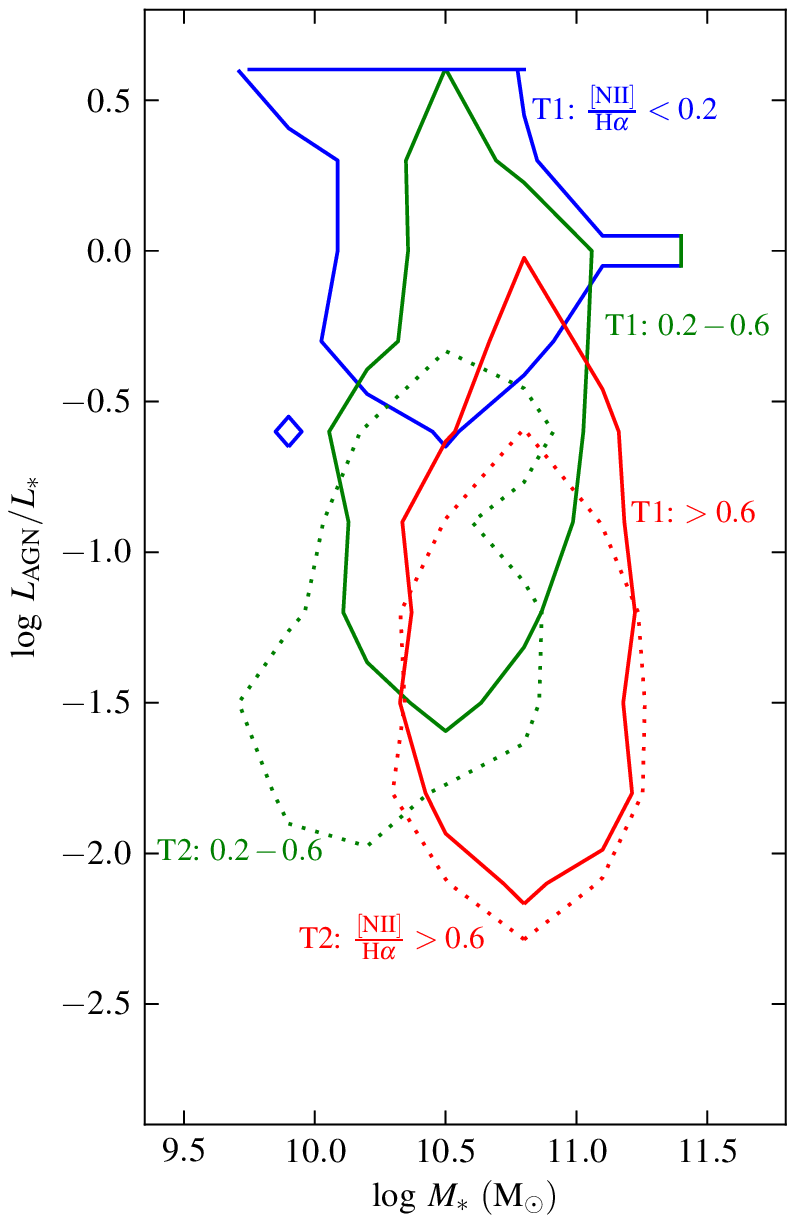}
\caption{
The distribution of T1 and T2 AGN samples with $\oiii/\Hb>5$ in the $M_*$ versus $\lagn/\lhost$ plane. 
Each sample is grouped into bins in \nii/\Ha. 
The contour line of each bin is equivalent to the inner contour of each group in Figure 6. 
The $\nii/\Ha$ range of each bin is noted near the contour, and coded by color. T1 contours are solid, and T2 contours are dotted. 
In the T1 sample, the distribution moves to lower $M_*$ and higher $\lagn/\lhost$ with decreasing $\nii/\Ha$. 
T2s follow a similar trend, but they do not extend to the high $\lagn/\lhost$ occupied by objects with $\nii/\Ha<0.2$ (Figure 6).
Thus, the absence of $\nii/\Ha<0.2$ AGN from the T2 sample just reflects its selection
against high $\lagn/\lhost$ objects, which are observed to have a low $\nii/\Ha<0.2$.
}
\label{fig: }
\end{figure}

A difference between type 1 and type 2 AGN can be either a failure of the unified model, an orientation-related effect, or simply due to different selection criteria used for creating the two samples. Here, we compare the T1 and T2 (\S2.4) samples, and show that selection effects are likely behind the differences observed in Figs. 2--5. 
To avoid significant NLR contamination by star formation in the host, which decreases the narrow line ratios to the Composite and SF regions of the BPT plots, we only use the 1\,691 T1s and 4\,042 T2s with $\oiii/\Hb > 5$. This selection criterion is independent of the offset quantities, \nii/\Ha\ and \sii/\Ha. A comparison of the Composites and SFs is made in \S8 below.

A major difference in the T1 versus T2 selection criteria, is that the T1 sample includes also point sources, and is not selected purely from extended objects. Thus, the T1 sample can extend to $\lagn/\lhost$ values larger than possible in the T2 sample. Another related systematic difference
is the distribution of $M_*$ values, as the T2 objects are selected by $\lhost$, while in
the T1 point sources $\lhost$ can be arbitrarily small. The distribution of $\lhost$ values is
interesting as $M_*$ was found to correlate with the \nii/\Ha\ ratio, via the $M_*-Z$ relation of galaxies, and the dependence of \nii/\Ha\ on $Z$ (Groves et al. 2006, see below). 

In Figure 6, we therefore plot contours of the distribution of T1s and T2s in the $M_*$ vs. $\lagn/\lhost$ plane. The T1 sample is divided according to the two SDSS surveys from which it is derived, those selected from the SDSS galaxy survey, and the point-sources from the SDSS quasar survey. 
We note that at $\lagn/\lhost > 1$, $\lhost$ (and $M_*$) can be significantly overestimated (\S2.3.1), therefore the true $\lagn/\lhost$ may be higher and the true $M_*$ may be lower than plotted. 
The abrupt cut at $\lagn/\lhost=3$ is due to the limit of our capability to derive a robust upper limit on $\lhost$ (\S 2.3.1). This limit does not affect the conclusions below. 
In the T2 sample, $\lagn$ is derived from $\loiii$ (Paper II), $\lhost$ is derived from the observed SDSS $z$-band luminosity, and $M_*$ is taken from Kauffman et al. (2003b). Note that by construction, we use the same $M/L$ in T1s and T2s (\S2.3.1).

The T2s are all selected from the SDSS galaxy survey. The distribution of the T1s from the same survey overlaps well the distribution of T2s. The T1 point sources however are clearly offset to higher $\lagn/\lhost$, and constitute an AGN population which does not appear in the T2 sample. 
As Figure 5 shows, the T1s become offset from T2s at the higher $\lagn/\lhost$ values. Thus, the 
apparent differences between the T1 and T2 BPT positions reflects their
different $\lagn/\lhost$ values, which controls the BPT positions.
Figure 7 explores this effect more quantitatively. The T1s and T2s are binned by the \nii/\Ha\ values, and the distributions of the different bins in the $M_*$ vs. $\lagn/\lhost$ plane is presented.  
In the T1 sample, $M_*$ decreases and $\lagn/\lhost$ increases with decreasing \nii/\Ha. 
The T2 sample shows a similar trend, but it does not extend to the high $\lagn/\lhost$ occupied by objects with $\nii/\Ha<0.2$. 
In fact, already T2s with $\nii/\Ha < 0.6$ are rare, constituting only 9\% of the T2 sample, compared to the 39\% of T1s that have $\nii/\Ha < 0.6$. 
Therefore, the T2 objects do not extend to the low \nii/\Ha\ values, seen in the T1 sample, as 
these values occur at high $\lagn/\lhost$ values, which the T2 objects cannot have by their selection.

A similar analysis using the \sii/\Ha\ values, instead of \nii/\Ha, demonstrates that the
low \sii/\Ha\ values seen in the T1 sample at high $\lagn/\lhost$ (Figure 5) are absent from the T2 sample for the same reason.

\begin{figure}
\includegraphics{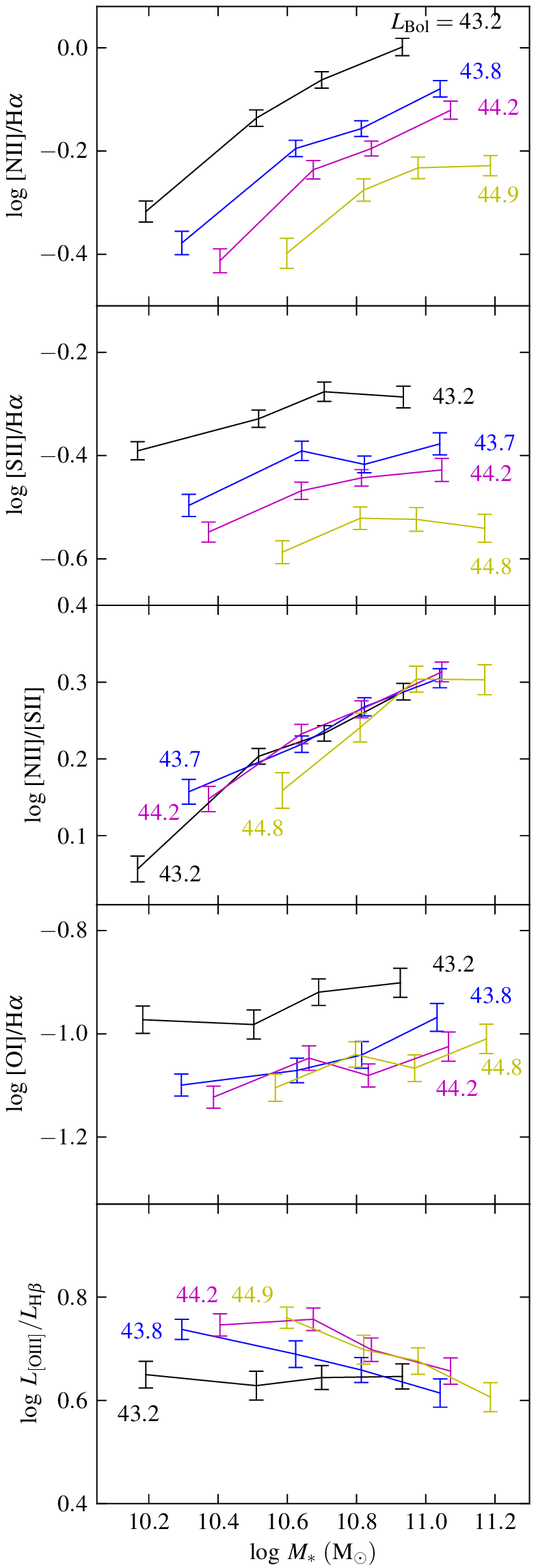}
\end{figure}
\begin{figure}
\caption{The dependence of the mean line ratios of the T1 objects on $\lbol$ and $M_*$. Given $\lbol$ bins are connected by solid lines (mean $\log\ \lbol$ noted). Error bars denote the uncertainty in the mean value. 
The 9\% of the T1 objects with inaccurate $M_*$ measurements ($\lagn/\lhost>1$), and 20\% of the T1s with significant host contamination of the NLR ($\luv/\lbha>100$) are not shown. 
{\bf Top panel} 
The mean $\nii/\Ha$ increases with $M_*$ at a fixed $\lbol$, 
and decreases with increasing $\lbol$ at a fixed $M_*$. 
{\bf Second panel} 
The mean $\sii/\Ha$ decreases with increasing $\lbol$ at a fixed $M_*$. 
The mean $\sii/\Ha$ increases only slightly with $M_*$ at a fixed $\lbol$. 
{\bf Third panel} 
The mean $\nii/\sii$ is determined by $M_*$ and is almost independent of $\lbol$. 
{\bf Fourth panel} 
The mean \oi/\Ha\ increases slightly with $M_*$ at a fixed $\lbol$, similar to the equivalent trend of \sii/\Ha. 
A dependence of $\oi/\Ha$ on $\lbol$ is seen only at the lowest $\lbol$ bin. 
{\bf Bottom panel}
In the three high $\lbol$ bins, $\oiii/\Hb$ decreases with $M_*$. 
The trends with $M_*$ likely reflect an increase of $\znlr$ with $M_*$. 
The trends with $\lbol$ is likely related to the decrease in NLR covering factor with increasing $\lbol$ (Paper II). 
}
\end{figure}

\section{Physical parameters of the NLR}

We now quantify the dependence of narrow line ratios on the observed AGN and host properties, and discuss the physical origin of the trends. 
To avoid a significant contribution to the NLR from star formation in the host galaxy, we require $\luv / \lbha < 100$ (\S8), instead of the $\oiii/\Hb>5$ requirement used above. This alternative cut is possible since we do not analyze T2s in this section, and enables us also to derive trends in $\oiii/\Hb$.

As can be seen in Figs. 3, 5 and 7, \nii/\Ha\ decreases both with increasing $\lagn$ and with decreasing $\lhost$. Below we explore the two effects independently.
We bin the T1 objects based on $M_*$ and $\lbol\ (\equiv 130\times\lbha)$\ in the following manner. The objects are sorted by $\lbol$ and divided into four equal size groups. Each of these groups is then sorted by $M_*$, and again divided into four equal size groups. This ensures similar statistical errors in all bins. We disregard the 12\% of the objects with $\lagn/\lhost>1$, in which the $M_*$ measured has a large error (\S2.3.1). 

Figure 8 presents the derived relations of the mean values of $\nii/\Ha$, $\sii/\Ha$, $\nii/\sii$, $\oi/\Ha$, and $\oiii/\Hb$ as a function of $M_*$, for different $\lbol$. Error bars denote the error in the mean. 
The upper panel shows that the mean $\nii/\Ha$ increases with $M_*$ at a fixed $\lbol$, an increase of $\sim 0.3$ dex over $\sim 0.7$ dex in $M_*$. Also, the mean $\nii/\Ha$ decreases with $\lbol$, at a fixed $M_*$, a decrease of $\sim 0.3$ dex over $\sim 1.7$ decades in $\lbol$. The trend of \nii/\Ha\ vs. $\lbol$ can also be seen in mean spectra, shown in the appendix.
The second panel shows that the mean $\sii/\Ha$ decreases with $\lbol$ at a fixed $M_*$, similar to the decrease in $\nii/\Ha$ with $\lbol$ in the upper panel. However, at a fixed $\lbol$, $\sii/\Ha$ increases only by $\sim0.1$ dex over $\sim 0.7$ dex in $M_*$. This small increase is within the range of possible systematics (\S 2.4), and in contrast with the sharper change in \nii/\Ha\ with $M_*$ in the top panel. 
The relative trends of \nii\ and \sii\ are most apparent in the third panel. Clearly, \nii/\sii\ strongly increases with $M_*$, and is almost independent of $\lbol$.
The fourth panel shows that $\oi/\Ha$ increases slightly with $M_*$ at a fixed $\lbol$, similar to the \sii/\Ha\ trend in the second panel. 
In the bottom panel, at $\log\ \lbol \geq 43.8$, $\oiii/\Hb$ shows a decrease of $\sim 0.15$ dex over $\sim 0.8$ dex in $M_*$.

What are the physical mechanisms behind these trends in narrow line ratios? 
The similarity of the behavior of \nii/\Ha\ and \sii/\Ha\ vs. $\lbol$, in contrast to the different behavior vs. $M_*$, suggests there are two distinct mechanisms at play. 
We address them separately below.

\subsection{The trend with $M_*$}

\subsubsection{$M_*$ vs. $\znlr$}

An increase of \nii/\Ha\ with $M_*$ has been observed in type 2 AGN by Groves et al. (2006), qualitatively similar to the trend we see in T1 AGN (Figure 8). As mentioned above, Groves et al. suggested this trend originates from the $M_* - Z$ relation found in quiescent galaxies. 
A relatively strong dependence of \nii/\Ha\ on $\znlr$ is expected since Nitrogen is a secondary nucleosynthesis product, and hence its abundance increases as $Z^2$ for $Z > 0.5\ Z_\odot$ (e.g. van Zee et al. 1998). Appropriately, an increase is expected also in the relative abundance of N to S, consistent with the increase of \nii/\sii\ vs. $M_*$ seen in Figure 8. Also, since $\oiii$ is a main coolant, the lower NLR temperature associated with the higher $\znlr$ is expected to reduce $\oiii/\Hb$, as observed in the bottom panel of Figure 8. The mild increase of $\sii/\Ha$ and $\oi/\Ha$ with $M_*$ are also consistent with an increase in $\znlr$ with $M_*$, if $\sii$ and $\oi$ are both trace coolants. 

Is the $\znlr$-based explanation of the trends vs. $M_*$ unique, or can these trends be explained by density / ionization effects? 
Density is an unlikely candidate, as the critical density $\ncrit$ of \nii\ is $10^{4.9}\ {\rm cm}^{-3}$, intermediate between $\ncrit(\sii)=10^{3.2-3.6}\ {\rm cm}^{-3}$ and $\ncrit(\oi)=10^{6.3}\ {\rm cm}^{-3}$ (all $\ncrit$ are taken from Appenzeller and {\" O}streicher 1988). Thus, if a change in the distribution of NLR gas densities is behind the trends vs. $M_*$, then the slope of the $\nii/\Ha$ vs. $M_*$ relation is expected to be intermediate between the slopes of the \sii/\Ha\ vs. $M_*$ and \oi/\Ha\ vs. $M_*$ relations, in contrast with Figure 8. 
For example, in an NLR model where the typical density decreases with increasing radius, the amount of obscuration of the dense inner region will affect the distribution of observed NLR densities (e.g. Zhang et al. 2008). 
In this scenario, the visibility of \oi-emitting clouds would be more sensitive to the amount of obscuration than the visibility of \nii-emitting clouds, which in turn would be more sensitive than \sii-emitting clouds. If obscuration decreases with $M_*$, one would expect \nii/\sii\ to increase with $M_*$, as observed in Figure 8, but one would also expect a steep slope of the \oi/\Ha\ vs. $M_*$ relation, which is not observed. Therefore, the trends vs. $M_*$ are unlikely to be related to the NLR density. 
Moreover, we find that the \sii\ doublet ratio ($\lambda6716$ to $\lambda6731$), which is sensitive to the density of the \sii-emitting gas, shows no dependence on $M_*$ in the T1 sample (absolute Pearson coefficient $<0.06$ for all luminosity bins). 

A similar argument can be used for ionization effects. If the NLR ionization changes with $M_*$, we would expect the $\nii/\Ha$ trend with $M_*$ to be intermediate between the trends of $\oi/\Ha$ and $\oiii/\Hb$, in contrast to the relative strength of the trends observed in Figure 8. 
However, given the flexibility in the current NLR models (e.g. Groves et al. 2004), one may be able to tune the NLR parameters and the change of ionization parameter with $M_*$ to reproduce the observed relations. 
Nevertheless, since an increase of $\znlr$ with $M_*$ explains the line ratio trends qualitatively from first principles, and since $Z$ is known to increase with $M_*$ in quiescent galaxies, and in type 2 AGN, a $\znlr$-based explanation for these trends appears more plausible. In the next section we provide additional support for this conclusion by showing that the BLR metallicity $\zblr$ also appears to increase with $\nii/\Ha$ at a fixed $\lbol$.

\subsubsection{$\znlr$ vs. $\zblr$}

Are there any additional differences in the spectra of objects with high and low $\nii/\Ha$?
Figure 9 compares the mean spectra of objects with $\nii/\Ha<0.2$ and objects with $\nii/\Ha>0.6$. 
To avoid other known trends, and isolate only $\nii/\Ha$ related trends, we match each of the T1s with $\nii/\Ha<0.2$ with a T1 that has $\nii/\Ha>0.6$ with the same $\lbol$ up to 0.1 dex, and the same $\dv$ up to 0.05 dex.
Matching by $\lbol$ ensures we are freezing the $\lbol$-related effect seen in Figure 8, 
while matching also by $\dv$ indicates we are freezing also $\mbh$ (via eq. 2 in Paper I) and $\lledd$
and the host of spectral properties related to it (e.g. Boroson \& Green 1992). 
Of the 184 T1s with $\nii/\Ha<0.2$, 148 have such matches.

The mean spectra of the two groups of objects are calculated by geometrically averaging luminosity densities of spectrum pixels with the same restframe wavelength $\lambda$, rounded to $10^{-4}$ in $\log \lambda$. The bottom spectrum is the difference between the two composite spectra, and the insets zoom in on the areas delimited by the dashed lines. The most striking feature of the residual is the strong BLR \feii\ multiplets at $\sim 4600\AA$ and $\sim 5300\AA$. 

\begin{figure*}
\includegraphics{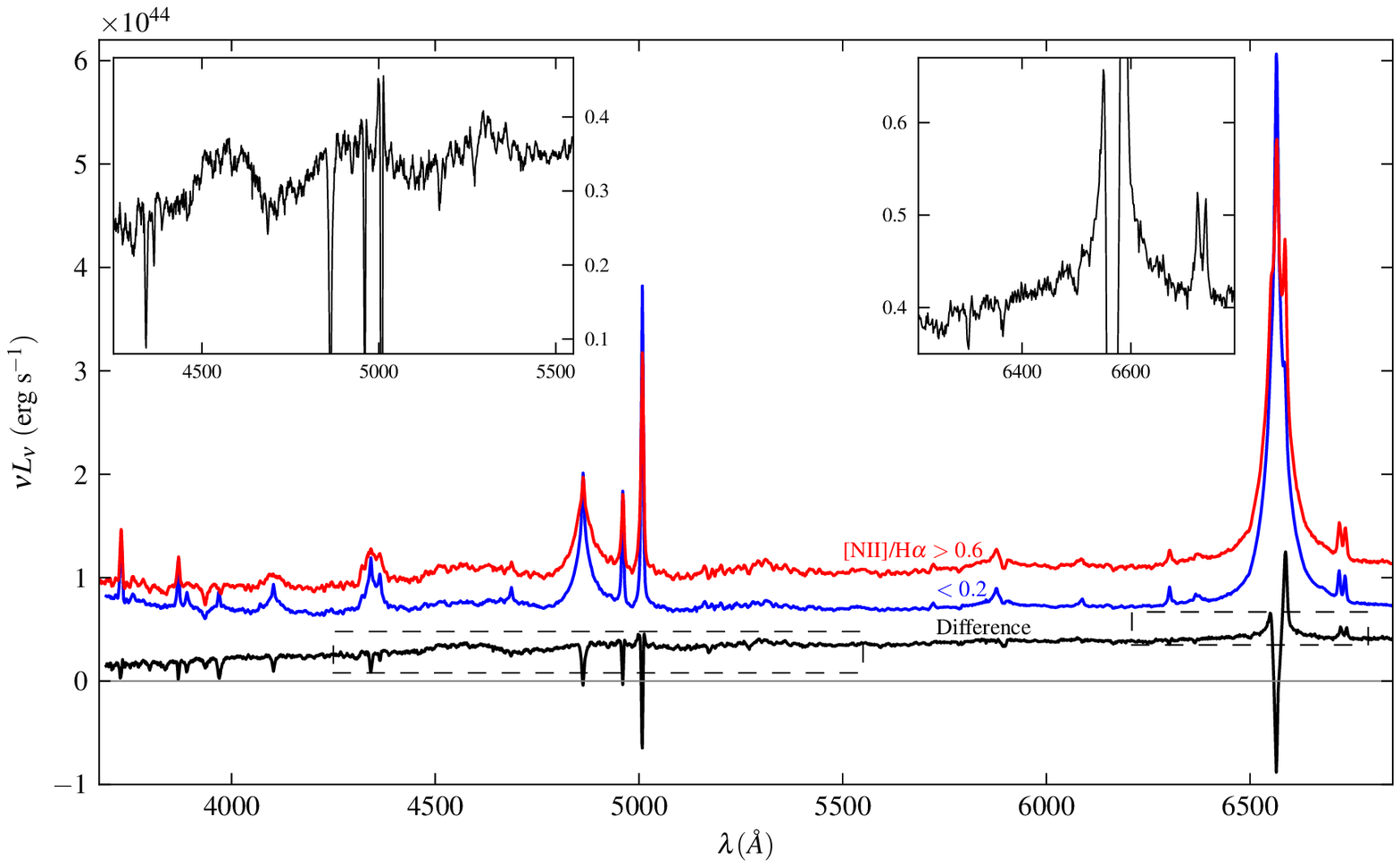}
\caption{
The difference between mean spectra of T1 AGN 
with high ($>0.6$) and low ($<0.2$) \nii/\Ha\ values. 
The two samples are selected to match in $\lbha$ and $\dv$, and thus in $\lledd$ as well,
Insets zoom in on the dashed rectangles in the difference spectrum.
The high \nii/\Ha\ spectrum also has stronger BLR \feii\ multiplets at $\sim 4600\AA$ and $\sim 5300\AA$, which are expected to increase with $\zblr$. Therefore, at a fixed $\lbol$ and $\lledd$, $\znlr$ and $\zblr$ appear to be related. 
The high \nii/\Ha\ spectrum also shows a larger $M_*$, evident from the red slope and the stellar absorption features at $3934\AA$ and $6500\AA$ in the residual. Thus, $\zblr$ also appears to be
at least partly driven by the host $M_*$.
}
\label{fig:  }
\end{figure*}

The luminosity of the optical \feii\ multiplets is expected to increase with iron column density, and therefore with $\zblr$, to a power of 0.8--0.9 (Verner et al. 2003, Baldwin et al. 2004, Shields et al. 2010). Thus, Figure 9 provides interesting evidence that $\zblr$ is related to 
$\znlr$. There is a  well-known relation between the \feii\ equivalent width and $\lledd$ (Boroson \& Green 1992), but since the two composites are matched in $\lledd$, this effect should not be present.

Shields et al. (2010) found that when binning by $L(\feii)/L_{\rm b\Hb}$, \nii/\sii\ increases by a factor of two for an increase of a factor of ten in $L(\feii)/L_{\rm b\Hb}$. They concluded that the \feii\ strength increases with $\znlr$, but the dispersion in \feii\ is not dominated by $\znlr$.
In Figure 9, the composite spectra differ by a factor of $2.3$ in \nii/\sii, implying a factor of 1.5 in $\znlr$ (see eq. 2 below). They also differ by a factor $\sim 2$ in $L(\feii)$. Therefore, for a constant $\lledd$, $\znlr$ and $\zblr$ change roughly in unison. 

The mean $\log\ M_*$ of the low and high \nii/\Ha\ composite spectra are 10.5 and 10.8, respectively. This difference in $M_*$ can be seen in the residual spectrum, which has a red optical slope, a \caii\ K $\lambda 3934$ absorption feature, a stellar absorption blend at 6500\AA, and at a few additional stellar features. The two groups are
selected to have the same mean $\mbh$, and should thus have similar mean bulge mass (Magorrian et al 1998). The different measured mean $M_*$ values of the two groups should therefore reflect differences in the mean disk masses, where the higher metallicity group has a higher disk/bulge mass ratio.

Hamann \& Ferland (1993, 1999) found that in quasars, $\zblr$ (derived from the NV / CIV ratio) increases with $\lbol$. They speculated that the increase in $\zblr$ with $\lbol$ is probably due to the increase of $\zblr$ with $M_*$, and the strong relation between $M_*$ and $\lbol$ in the quasar samples they used, where most objects
shine close to the Eddington limit. Their conclusion is supported by the increase of NV / CIV with $\mbh$, which should also increase with increasing $M_*$ (Warner et al. 2003).
Here, we confirm their claim by showing that $\zblr$ increases with $M_*$ directly. 

At a given $M_*$, the mean $\nii/\sii$ remains constant with $\lbol$ (Figure 8). 
Therefore, we find no evidence for a direct $Z - \lbol$ trend.

\subsubsection{Estimating ${\rm O/H}$ from $\nii/\sii$}

Since the NLR is the part of the ISM which is located on 10s -- 100s pc from the nucleus and is exposed to the ionizing AGN radiation, it is plausible that $\znlr$ is the gas phase $Z$ of the host. 
Therefore, given a calibration between $\nii/\sii$ and the gas phase absolute metallicity, as indicated by the
oxygen abundance ${\rm O/H}$, we can use $\nii/\sii$ to estimate ${\rm O/H}$ in the host galaxy. 

In principle, we could apply the ${\rm O/H}$ vs. $\nii/\sii$ relation of \hii\ regions to the NLR. However, the different physical conditions in the NLR and \hii\ regions of star forming galaxies may imply that the NLR has a different ${\rm O/H}$ vs. $\nii/\sii$ relation. 
Instead, we use the relation of $\nii/\sii$ vs. $M_*$ in the T1 sample and the ${\rm O/H}$ vs. $M_*$ relation from Tremonti et al. (2004, hereafter T04) to indirectly calibrate ${\rm O/H}$ vs. $\nii/\sii$ in the NLR. 

T04 found that the median ${\rm O/H}$ in SDSS star forming galaxies follows $12 + \log ({\rm O/H}) = -0.08 m_{10}^2 + 0.25 m_{10} + 9$ for $-1.5 < m_{10} < 1.5$, where $m_{10} = \log (M_*/10^{10}\ \msun)$. 
Similarly, we fit a 2\nd-order polynomial relation to the median $\nii/\sii$ vs. $M_*$ relation in the T1 sample. Using all 0.1-dex bins in $M_*$ with $>10$ objects, we find 
\begin{equation}
\log \nii/\sii = -0.17 m_{10}^2 + 0.46 m_{10} - 0.01
\end{equation}
for $0 < m_{10} < 1.3$. The typical dispersion in each $m_{10}$ bin is 0.15 dex. 

Eq. 1 shows a flattening of the $\nii/\sii$ vs. $m_{10}$ relation with increasing $m_{10}$, similar to the flattening of the T04 $m_{10}$ vs. $Z$ relation. This similarity supports the suggestion that the $\znlr$ is the host gas phase $Z$. Plugging eq. 1 in the T04 relation we get
\begin{equation}
12 + \log\ {\rm O/H} = 0.47\ \log\ \nii/\sii + 9.03\ \ \  (\sigma \sim 0.06\ {\rm dex})
\end{equation}
where we neglected a term equal to $0.03m_{10}$ on the right hand side. The dispersion $\sigma$ in eq. 2 is the dispersion of the $\nii/\sii$ vs. ${\rm O/H}$ relation in the T04 star forming galaxies, which could be biased due to the different physical conditions in \hii\ regions and in the NLR.

For comparison, in the T04 star forming galaxies we find $12 + \log\ {\rm O/H} = 0.73\ \log\ \nii/\sii + 8.94$. 
Eq. 2 can be used as a rough estimate for ${\rm O/H}$ in AGN hosts.

\subsection{The trend with $\lbol$}

What is the source of the change in \nii/\Ha\ and \sii/\Ha\ with $\lbol$? 
In Paper II, we found that $\lnha/\lbha$ decreases with $\lbol$, and presented evidence that this trend is due to a decrease in the NLR covering factor ($\cfnlr$) with $\lbol$. We verify this trend depends on $\lbol$ and not on $M_*$, by measuring $\lnha/\lbha$ vs. $M_*$ at a given $\lbol$, using the same bins as shown in Figure 8. Indeed, in all $\lbol$ bins $\lnha/\lbha$ changes by $<0.1$ dex over $0.8$ dex in $M_*$. The $\lnha/\lbha$ ratio 
thus depends purely on $\lbol$.
Therefore, it seems that the decrease in \nii/\Ha\ and \sii/\Ha\ with $\lbol$, at a given $M_*$, is associated with the decrease in $\cfnlr$. 

A change in $\cfnlr$ alone cannot change the narrow line ratios. Therefore, the distribution of some other NLR physical parameter such as $Z$, density or ionization  probably also changes with $\lbol$. As mentioned above, a change of $\znlr$ with $\lbol$ is unlikely. We discriminate between a change in density and ionization using the Baldwin slopes (Figure 1). 
The Baldwin slopes $\alpha$ of the different lines follow $\alpha_\oiii > \alpha_\oi > \alpha_\nii \sim \alpha_\sii$. This order favors a change in the density distribution of the NLR over a change in the ionization distribution, since \oiii\ and \oi\ have higher $\ncrit$ than \nii\ and \sii, while the ionization energies of \nii\ and \sii\ are intermediate between the ionization energies of \oiii\ and \oi. Therefore, a possible scenario which explains the observed trends vs. $\lbol$ is that the covering factor of the clouds with density $10^3-10^4\ {\rm cm}^{-3}$ drops faster with increasing $\lbol$ than the covering factor of the clouds with density $10^5-10^6\ {\rm cm}^{-3}$.

We emphasize that since the trend with \nii/\Ha\ with $\lbol$ is probably not a $\znlr$ effect, deriving $\znlr$ in quasars from $\nii/\Ha$ calibrated on lower luminosity AGN (e.g. Husemann et al. 2011), will underestimate $\znlr$.

\section{LINERs}

Ke06 found that at a fixed $\lledd$, the difference between host properties of Seyferts and LINERs\footnote{It is disputed whether the narrow lines of SDSS LINER 2s with low \oiii\ equivalent width, and therefore low implied $\lledd$, are powered by AGN (e.g. Sarzi et al. 2010). This caveat does not affect our conclusions, therefore we disregard it in the following analysis.} disappear. 
Their conclusion was that the observed difference in host properties between Seyferts and LINERs is only a secondary effect, which results from their difference in $\lledd$ (Ho 2002, Ke06). Here, we show that
the observed large difference between Seyferts and LINERs in terms of the fraction which shows broad lines (Ho et al. 1997b, Ho 2008), is also a secondary effect of their difference in $\lledd$, and at a fixed $\lledd$ the difference disappears.

Following Ke06, we create subsamples of the T1 and T2 samples which include objects classified as AGN in the BPT-\nii\ panel, and as either Seyferts or LINERs in the BPT-\sii\ and BPT-\oi\ panels (Figure 2). We use only objects with consistent BPT-\sii\ and BPT-\oi\ classifications. 
We use the bulge stellar dispersion $\sigma_*$ to derive $\mbh$ in T2s (G{\"u}letkin et al. 2009). We disregard the 8\% of the T2s with surface mass density $< 3 \times 10^8\ \msun\ {\rm kpc}^{-2}$, in which the $\sigma_*$ measured by the SDSS may be overestimated due to disk light contamination (Kauffmann et al. 2003c, Heckman et al. 2004). 

Following the above criteria, the T2 subsample includes 4\,938 Seyfert 2s and 4\,292 LINER 2s. The T1 subsample includes 1910 Seyfert 1s and 76 LINER 1s. Thus, LINERs constitute 50\% of the T2 sample, but
only 4\% of the T1 sample. Our purpose is to further understand the origin of this large difference.

In 44 objects from the LINER 1 group, the classification is ambiguous, either due to upper/lower limits on the BPT ratios, or because their narrow line ratios are poorly constrained (\S2.2.3). We address this uncertainty below. The fraction of Seyfert 1s with an ambiguous classification is negligible.

In the T1 sample, we derive $\mbh$ from $\lbha$ and $\dv$, using eq. 2 in Paper I. 
For $\lbol$, we use $\lbol = 130\ \lbha$ (eq. 6 in Paper I). 
In the T2 sample, we derive $\lbol$ from $\loiii$ using the $\loiii-\lbha$ relation in the T1 sample (eq. 3 in Paper II), and the same $\lbol/\lbha$ as for the T1 sample. 
We note in passing that the $\loiii/\lbha$ ratio is expected to be lower in LINERs, almost by their definition (see factor of two drop in $\loiii/\lbha$ in the lower-left panel of Figure 6 in Paper II). So, due to this effect, the implied $\lbol$ in LINER 2s may be a bit underestimated. 
Additionally, LINERs might have a different $\lbol/\lbha$ ratio than the ratio we use, as this ratio was derived on the T1 sample, which is dominated by Seyferts. However, this latter caveat will affect our estimate of $\lbol$ in LINER 1s and LINER 2s in the same way, and will therefore not affect our analysis.

Figure 10 presents the fraction of LINERs 2 out of the T2 sample, as a function of $\lledd$. Seyfert 2s and LINER 2s are cleanly separated in $\lledd$, as found by Ke06. At $\lledd > -2$ all T2s are Seyferts, while at $\log \lledd < -4$ all T2s are LINERs. 
A similar clean cut in NLR ionization level can be seen in radio galaxies, where most $\lledd > 10^{-3}$ objects have $\oiii/\oii\ \lambda 3727 > 1$, while all $\lledd<10^{-3}$ objects have $\oiii/\oii\ < 1$ (fig. 9 in Antonucci 2012, Ogle et al. in prep.). 
Figure 10 also presents LINER 1 fractions out of the T1 sample, at different $\lledd$. The uncertainty in the LINER 1 fraction is due to the 44 T1 objects with an ambiguous LINER classification. 
At a fixed $\lledd$, the fraction of LINER 1s is consistent with the fraction of LINER 2s. Therefore, 
the small fraction of LINERs in the T1 sample results from the fact that the sample does not
extend to low enough $\lledd$, where LINERs become the dominant population.

\begin{figure}
\includegraphics{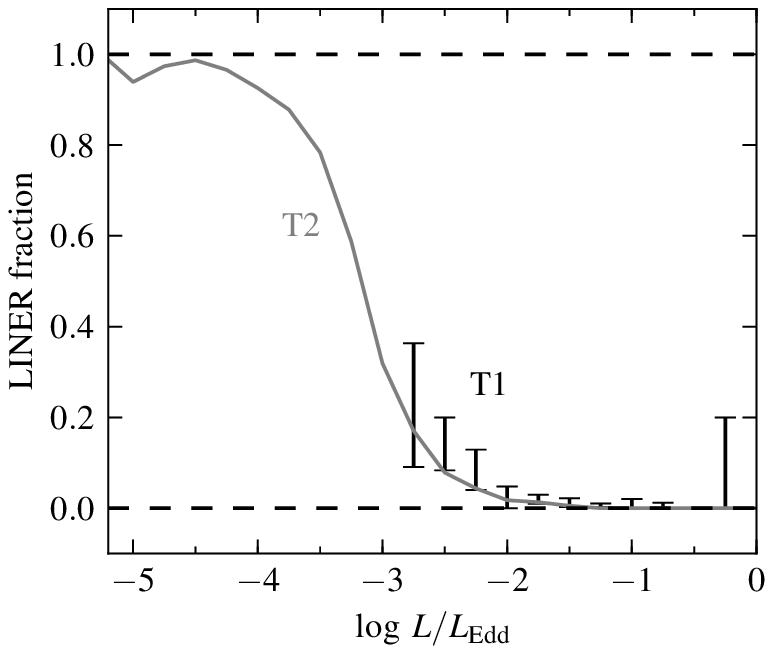}
\caption{
The fraction of LINERs in  the T1 and T2 samples, as a function of $\lledd$. The solid gray line indicates the fraction of LINERs in the T2 sample, in 0.25 dec $\lledd$ bins, where $\lledd$ is derived from $\loiii$ and $\sigma_*$.
The fractions of LINERs in the T1 sample are denoted by error bars, where $\lledd$ is derived from $\lbha$ and $\dv$. The uncertainty is due to T1 objects with an ambiguous classification. 
At $\log\ \lledd>-2$ all T2s are Seyferts, while at $\log \lledd < -4$ all T2s are LINERs. 
At a fixed $\lledd$, the fraction of LINER 1s is consistent with the fraction of LINER 2s within the uncertainties. 
The low fraction of LINERs in the T1 sample ($\sim 4\%$) versus the high fraction in the T2 sample
($\sim 50\%$) results from the difference in the $\lledd$ distribution of the T1 and T2 samples. The lack of $\lledd < 10^{-3}$ T1s could be due to detection limits, or due to a physical absence of low $\lledd$ type 1 AGN.
}
\label{fig: }
\end{figure}

Why do the T1 and T2 samples differ in their $\lledd$ distribution? 
The fact that all T1s have $\lledd > 10^{-3}$ could be a detection limit, since low $\lledd$ have weak and wide broad \Ha\ features, which are hard to distinguish from the stellar continuum (see Figure 6 in Paper I). Alternatively, there may be a physical reason for a lack of low $\lledd$ type 1 AGN,
related to the subject of `true type 2' AGN
(see further discussions in Laor 2003; Laor \& Davis 2011). 

The fact that the probability an object is a Seyfert or a LINER does not depend on whether
the BLR is detected or not, indicates the transition from Seyferts to LINER does not affect the
BLR. This may indicate that the physical difference between these two type of objects occurs beyond the BLR, and hence external to the central source. I.e., Seyferts and LINERs may differ by the conditions in the circumnuclear gas, and not by a different accretion mode, as suggested by Dudik et al. (2009). Such a scenario implies that the intrinsic UV and X-ray emission of LINERs and Seyferts should not be distinct, as found by Maoz et al. (2005, 2007). Though, these latter results are disputed (see review by Ho 2008).

\section{T1 AGN classified as Composites and SF}

Why do some of the T1 objects display narrow line ratios characteristic of Composites and SF galaxies?
Can such line ratios be powered by accretion onto a massive black hole, or does it result from
host contamination?
The fraction of T1s classified as Composites increases with $\lhost/\lagn$ at the SDSS-$z$ band (Figure 5),
which suggests a host contamination effect. Below we explore quantitatively the host contamination, based on other indicators, and its relation to the narrow line ratios. We compare the $\lnha/\lbha$ and $\luv/\lbha$ of Composites with those of T1s which fall above the Ke01 line in the BPT-\nii\ panel (hereby called `pure-AGN'). In pure-AGN $\lnha$ and $\luv$ correlate with $\lbha$, 
thus host contribution should manifest as higher $\lnha/\lbha$ and $\luv/\lbha$ due to line and continuum emission from the SF regions. 

In Table 4, we list the geometrical mean of $\lbha$ and $\lnha/\lbha$ for the T1 AGN classified as pure-AGN, Composites and SF. 
The SF group is divided into `SF-robust' (32 objects) and `SF-non robust' (69 objects), depending on whether their narrow line ratios are well-constrained (\S 2.2.3). This division is 
to guard against systematic uncertainties in the less secure measurements.
As seen in the lower left panel of Figure 4, non-robust SFs tend to have high $\lledd$, where the NLR is weak and the broad \Ha\ is relatively narrow, making the NLR / BLR deblending difficult. It is therefore possible that in non-robust SFs broad Balmer flux was mistakingly assigned to the narrow Balmer lines, and their SF classification is not real. 
In the Composite and pure-AGN classes poorly constrained objects are less abundant (17\%), and therefore a separate group is not required. 

\begin{table}
\begin{tabular}{l|c|c|c|c}
BPT-\nii\ Classification & $N$ & $\lbha$ & $\lnha/\lbha$ & $\luv/\lbha$  \\
\hline
\hline
            pure-AGN  &  2303  &  42.0  &  $0.11$  &  $33$  \\
\hline
          Composites  &   407  &  41.8  &  $0.27$  &  $54$  \\
~~~~~$L$ matched pure-AGN  &   407  &  41.8  &  $0.12$  &  $36$  \\
\hline
           SF-robust  &    32  &  41.6  &  $0.44\pm 20\%$  &  $62\pm 20\% $  \\
~~~~~$L$ matched pure-AGN  &   128  &  41.6  &  $0.14\pm 10\%$  &  $40\pm 10\%$  \\
\hline
       SF-non robust  &    69  &  42.8  &  $0.16\pm 10\%$  &  $37\pm 10\%$  \\
~~~~~$L$ matched pure-AGN  &   207  &  42.7  &  $0.06\pm 10\%$  &  $29\pm 10\%$  \\
\end{tabular}
\caption{The mean $\lnha/\lbha$ and $\luv/\lbha$ of different BPT-\nii\ classifications. The SF objects are divided according to whether their narrow line measurements are robust (\S2.2.3). In the pure-AGN and composite groups, 83\% of the objects have robust measurements. The $\lbha$ (in $\log\ \ergs$), $\lnha/\lbha$ and $\luv/\lbha$ values are the geometrical means, with the uncertainty in the mean noted only if it is $>5\%$.
}
\end{table}

Since the mean AGN $\lnha/\lbha$ decreases with increasing AGN luminosity (Paper II), we compare each classification with a pure-AGN matched in $\lbha$. The matched groups are constructed by randomly selecting 1--4 pure-AGN T1 objects with the same $\lbha$ (up to 0.1 dex), for each Composite or SF (see Table 4).
The geometrical mean $\lnha/\lbha$ of the Composites is 0.27, compared to 0.12 in the matched pure-AGN. 
Therefore, the $\lnha/\lbha$ ratios of Composites are consistent with a roughly equal AGN and host contribution to $\lnha$. In the robust SFs, the host contribution is twice the AGN contribution. An intermediate ratio is seen in the non-robust SFs.

A similar effect is expected in $\luv/\lbha$, as star formation will contribute only to $\luv$. 
Indeed, the mean $\luv/\lbha$ of Composites and robust SFs is 50\% higher than in the respective matched group 
(25\% difference in the non robust SFs). Is the observed increase in $\luv/\lbha$ consistent with the observed increase in $\lnha/\lbha$? 
Star forming galaxies have a mean $\luv/\lnha = 120$ (Kennicutt \& Evans 2012). 
The Composites show an increase of 0.15 in $\lnha/\lbha$, and are thus expected to show
an increase of $120\times 0.15=18$ in $\luv/\lbha$, which is indeed observed (54 from 36, Table 4).
The robust SF group show an increase of 0.3 in $\lnha/\lbha$, and are thus expected to show
an increase of 36 in $\luv/\lbha$, which is 50\% larger compared to the observed rise of 22.
However, the difference is probably consistent within the larger uncertainties in this group. 
In the non robust
SF group the expected rise in $\luv/\lbha$ is 12, versus an observed value of 9, again 
consistent with the uncertainties. 

To summarize, the T1 AGN which reside in the Composites and SF regions of the BPT diagrams, also
show higher $\luv/\lbha$ and $\lnha/\lbha$ ratios, compared to pure-AGN. In addition, the ratio of
the increase in $\lnha$ and in $\luv$ is consistent with $\luv/\lnha$ observed in star forming galaxies.
Thus, AGN powered by accretion onto a massive BH do not produce SF or Composite line ratios, and
measurements of such line ratios in AGN implies host contamination.

Could host contamination also affect line ratios within the pure-AGN regime?
Could some of the spread in the BPT diagrams, also within the pure-AGN regime, be caused by host 
contamination? Figure 11 presents the mean BPT positions of T1s binned by $\luv/\lbha$. We split
the T1 sample to $\lbha<10^{42}\ \ergs$ (upper panels), and $\lbha>10^{42}\ \ergs$ (lower panels). The luminosity cut is set where
the host contribution to $\luv$ starts to be significant (Paper I). At $\lbha>10^{42}\ \ergs$, 
objects within the $\luv/\lbha\le 40$ bins have similar mean positions, but the highest bin
$\luv/\lbha=80$ is shifted towards the Composite region. A similar behaviour is observed 
at $\lbha<10^{42}\ \ergs$. 
Objects within the $\luv/\lbha\le 30$ bins have similar mean positions, but the $\luv/\lbha=90$ bin
is shifted towards the Composite region. The highest bin here has $\luv/\lbha=200$, and its mean position is within the Composite region. Thus, not only that Composite AGN have a higher mean $\luv/\lbha$, as found earlier, also the highest $\luv/\lbha$ AGN are on average composite in nature. 
Therefore, the excess UV, seen in low luminosity AGN, likely arises from star formation in the host, as suggested in Paper I, based on a comparison of their SED to the pure AGN SED.

In addition, AGN within the `pure-AGN' BPT regime can also be affected by host contamination, in particular when getting close to the Ke06 line. Narrow emission lines, powered purely by accretion,
likely produces a smaller dispersion than observed in the BPT plots. 

\begin{figure*}
\includegraphics{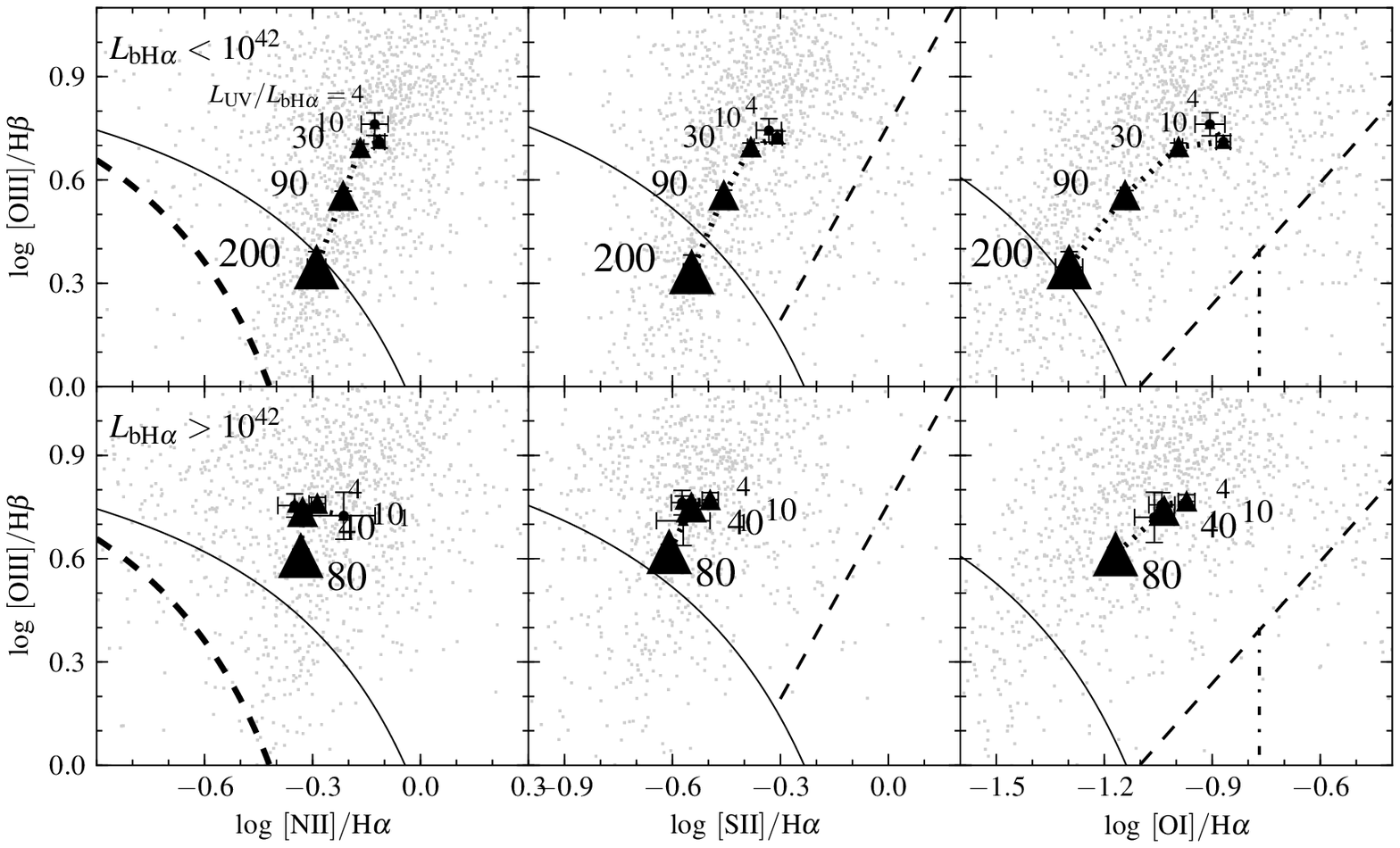}
\caption{The mean BPT positions by $\luv/\lbha$, for low and high luminosity T1s. Mean positions of 0.5 decade $\luv/\lbha$ bins are denoted by triangles, with marker size increasing with $\luv/\lbha$ (mean value noted). The error bars indicate the uncertainty in the mean position. Only bins with $>10$ objects are shown.  The positions of individual objects are shown as gray dots. The solid, dashed and dashed-dot classification lines are as in Figure 2. 
{\bf Top panels} 
T1s with $\log\ \lbha < 42$. The mean position of objects with $\luv/\lbha = 90$ and 200 are offset towards the composite region, indicating the excess UV originates from star formation in the host galaxy. 
{\bf Bottom panels} 
T1s with $\log\ \lbha > 42$. The $\luv/\lbha = 40, 10, 4, 1$ bins have similar mean BPT positions, indicating a similar intrinsic ionizing spectrum at these different $\luv/\lbha$. Therefore, if $\luv/\lbha<40$ indicates dust extinction (Paper I), the extincting dust resides on scales larger than the NLR.
}
\label{fig: }
\end{figure*}

\section{The ionizing spectrum seen by the NLR}

\subsection{$\aox$ as a measure of the ionizing spectrum slope}

What produces the scatter in the BPT plots? Possible parameters are the ionizing spectral slope and
the ionization parameter (e.g. Groves et al. 2004). Below we test this explanation by exploring
the dependence of the BPT positions on $\aox$, the power law slope interpolated from $\luv$ and $\lx$. 

We use the 752 T1 objects that were observed by GALEX and have $\lbha>10^{42.5}\ \ergs$, to avoid host contamination of the UV. 
We note that this luminosity cut limits the AGN luminosity dynamical range to $10^{44.5} < \lbol < 10^{46}\ \ergs$. 
These T1 objects are divided into bins of $\aox$ with width of 0.25. Figure 12 shows the mean BPT-\oi\ positions of the different $\aox$ bins. Error bars denote the error in the mean position. We use the BPT-\oi\ panel since it is most sensitive to the ionizing slope (Groves et al. 2004). 
The X-ray detection rates are 77\%, 73\%, 67\% and 29\%, for the $\aox=$ -1.2, -1.4, -1.6 and -1.8 bins, respectively. The UV detection rate is 60\% for the $\aox=-1.2$ bin, and $>95\%$ in the other bins. Upper limits are used when a detection is not available, so the true $\aox$ of the $\aox=-1.8$ bin is likely $<-1.8$, while the true $\aox$ of the $\aox=-1.2$ bin is likely $>-1.2$. 

\begin{figure}
\includegraphics{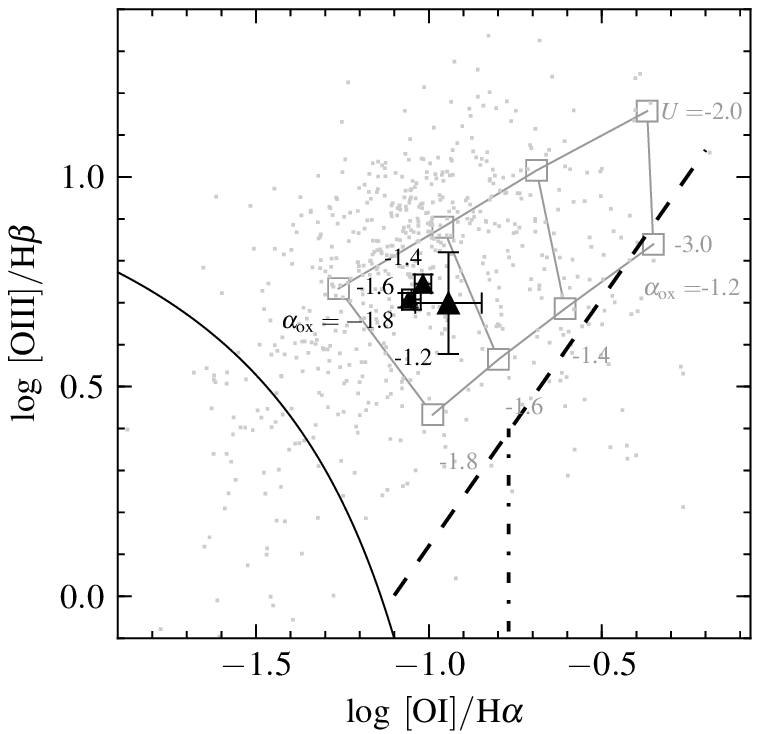} 
\caption{
The mean BPT--\oi\ positions of the T1 sample, for different $\aox$. The solid, dashed and dashed-dot classification lines are as in Figure 2. 
To avoid host contamination of $\luv$, only T1 objects with $\lbha>10^{42.5}\ \ergs$ are used (gray dots). 
The mean position of each $\aox$ bin is marked by a triangle (mean $\aox$ noted), with error bars denoting the error in the mean position. 
For comparison, the expected BPT position for ionizing spectra with different slopes and different $\log\ U$ are marked (values from Groves et al. 2004). 
The mean BPT positions of the T1 objects do not follow the trend expected if the slope of the ionizing spectrum is $\aox$, indicating that either $\aox$ does not represent the EUV spectral slope observed by the NLR, or other parameters, such as metallicity, ionization and density control the position. 
}
\label{fig: }
\end{figure}

We note that the known trend of $\aox$ vs. AGN luminosity (e.g. Just et al. 2007) implies a range of $0.2$ in the mean $\aox$ over the luminosity range spanned by the objects shown in Figure 12 (see fig. 20 in Paper I). Therefore, the observed range of $0.6$ in $\aox$ in these objects is not dominated by the global trend with AGN luminosity. 

For comparison, Figure 12 also shows the expected BPT-\oi\ position for ionizing spectra with different slopes and for different ionization parameters, taken from figure 1d in Groves et al. (2004), which assume a density of $1000\ {\rm cm}^{-3}$ and $Z = 2\ Z_\odot$. Clearly, the observed mean position is independent of the mean observed $\aox$, in sharp contrast with the models which predict a strong dependence.
This discrepancy may indicate that at a given luminosity, the spread in $\aox$ does not reflect a spread in the ionization slope at the EUV. The dispersion in the BPT plots is produced by another parameter, such as $Z$,
ionization parameter and the NLR density.

Telfer et al. (2002) showed that the mean EUV slope of $0.33<z<1.5$ quasars, observed by HST, is consistent with the mean $\aox$ of quasars with the same luminosity, confirming previous results by Laor et al. (1997). Therefore, the mean EUV slope and mean $\aox$ do seem to coincide. 
However, Figure 12 suggests that this equality does not extended to individual AGN.
There may exist additional mechanisms which produces a dispersion in $\aox$ with no effect on the BPT positions. For example, variability on timescales shorter than the NLR light crossing time ($\gtrsim 100$ yrs). However, Vagnetti et al. (2010) showed that variability on timescales of up to one year accounts only for 30 -- 40\% of the scatter in $\aox$ at a given AGN luminosity. 
Another source for a dispersion in $\aox$ is absorption restricted to our line of sight. A dusty absorber will flatten
$\aox$, as the dust optical absorption opacity is significantly larger than the X-ray absorption opacity
(e.g. Laor \& Draine 1993), while a dustless absorber will absorb only the X-ray and will steepen $\aox$, as commonly seen in broad absorption line quasars (e.g. Brandt et. al. 2000). An absorber restricted to our
line of sight will not significantly affect the NLR emission, and thus the BPT position will remain unchanged.
A third option is an absorber located outside the NLR, so the NLR sees the intrinsic ionizing spectrum, and the BPT ratios are not affected. In the next section we show that such a distant dusty absorber does exist in AGN.

\subsection{The effect of dust on the ionizing spectrum} 

Some AGN appear to be dust reddened based on their SED (e.g. Richards et al. 2003).
In Paper I, we found that the $\luv/\lbha$ distribution at the high luminosity end of the T1 sample is at least partially due to dust reddening along the line of sight. In particular, objects with $\luv/\lbha<30$ 
show a correlation such that redder optical slopes go with a decreasing $\luv/\lbha$.
A possible correlation between the reddening and $\lhost/\lagn$ suggested that this dust resides on host galaxy scales, beyond the NLR. 
This suggestion can now be tested using Figure 11, which shows the mean BPT positions by $\luv/\lbha$. The mean positions of the $\luv/\lbha = 40, 10, 4, 1$ bins are all similar to each other, in both high luminosity T1s (lower row) and low luminosity T1s (upper row). 
If the dust resides inside the NLR, then the NLR in objects with a low $\luv/\lbha$ is illuminated by a modified ionizing SED, which will shift their mean BPT position.
The complete lack of a trend in BPT position with reddening suggests that the NLR illumination is not
modified, and therefore the extincting dust resides on scales larger than the NLR.

\section{Conclusions}
The narrow line ratios of type 2 AGN have been extensively explored, in particular based on the SDSS sample. Here we present a similar analysis of the T1 sample, a large (3\,175 objects) sample of type 1 AGN (Paper I).
The T1 sample extends to luminosities well below the SDSS quasar sample, and thus in contrast
with quasars, where the narrow lines are generally difficult to measure, here a 
significant fraction of the objects have strong narrow lines (Paper II). This allows reliable
analysis of the narrow line ratios for most objects, as done in type 2 AGN.
We find the following:

\begin{enumerate}
\item The luminosities of all measured narrow lines, \Ha, \Hb, \oiii, \nii, \sii, \oi, show a Baldwin relation
relative to the broad \Ha\ luminosity, $L_{\rm line}\propto \lbha^{\alpha}$, with
$\alpha=0.66,\ 0.67,\ 0.72,\ 0.54,\ 0.53,\ 0.63$, respectively (Paper II, and above).

\item About 20\% of the T1 AGN have line ratios within the `Composite' and `SF' regions of the BPT diagrams. 
These line ratios are not powered by accretion onto a massive BH, as 
these objects also show higher $\lnha/\lbha$ and $\luv/\lbha$ emission. The excess $\lnha$ and
$\luv$ is consistent with the ratio expected from SF in the host galaxy, and indicates the line
emission in these objects is mostly excited by SF, rather than by the AGN.

\item The other 80\% of the T1 AGN, which reside within the BPT AGN region, are offset to lower \nii/\Ha\ and \sii/\Ha\ luminosity ratios, compared to type 2 AGN. This offset is a selection effect, as T1 AGN selected 
only from the SDSS galaxy sample, as the type 2 AGN are, are not offset. The offset is produced by
the T1 point like objects, selected from the SDSS quasar sample, which extend to higher $\lagn/\lhost$. 
The T2 sample is selected against such objects, and such objects are offset to lower \nii/\Ha\ and \sii/\Ha.

\item The \nii/\Ha\ and \nii/\sii\ ratios increase with host mass, which suggest a 
mass-metallicity relation in AGN hosts, as observed in quiescent galaxies. In contrast,
\nii/\Ha\ decreases with $\lagn$,  but \nii/\sii\ is independent of $\lagn$, which indicates there is no
direct $\lagn$-metallicity relation. 

\item At a fixed $\lbol$ and $\lledd$, objects with a higher \nii/\Ha\ also have higher
broad \feii\ luminosity, suggesting the broad line metallicity is also related to the host mass. This may be an
additional independent effect to the $\lledd$-metallicity relation, suggested in earlier studies to
explain some of the eigenvector 1 relations.

\item The fraction of AGN which are LINERs increases from $\sim 0$ at $\lledd=10^{-2}$ to
$\sim 1$ at $\lledd=10^{-4}$. The T1 and T2 samples show a similar fraction at a given $\lledd$,
indicating the LINER phenomena is unrelated to the presence of an observable BLR.
However, the T1 sample terminates at $\lledd\sim 10^{-3}$, either due to a physical effect or due to
selection effects, and thus LINERs constitute only $\sim 4$\% of the T1 sample, but $\sim 50$\% of the T2 sample.

\item The BPT position is unaffected by the value of $\luv/\lbha$ for values $<30$, which provide a measure 
of the foreground dust extinction (Paper I). This suggests that the ionizing continuum observed
at the NLR is unaffected by dust extinction, and the dust likely resides on the host galaxy
scale. 

\item The BPT position of $\lbol \sim 10^{45}\ \ergs$ AGN is unaffected by the observed spread in $\aox$. 
Models show there is a strong dependence of the BPT position on the ionizing continuum slope. This suggests that
the scatter in $\aox$ is dominated by mechanisms which do not affect the ionizing slope seen by the NLR,
such as absorption along our line of sight, or outside the NLR. Also, this result suggests that parameters other than the ionizing continuum slope, such as metallicity, density, and ionization parameter, 
dominate the scatter in the BPT plots.
\end{enumerate}

We thank Dan Maoz, Hagai Netzer, Minjin Kim, Luis Ho, and the anonymous referee for helpful suggestions and comments. 
We thank God for finishing this work...

\appendix
\appendixpage

\section{Mean spectra}

\begin{figure*}
\includegraphics{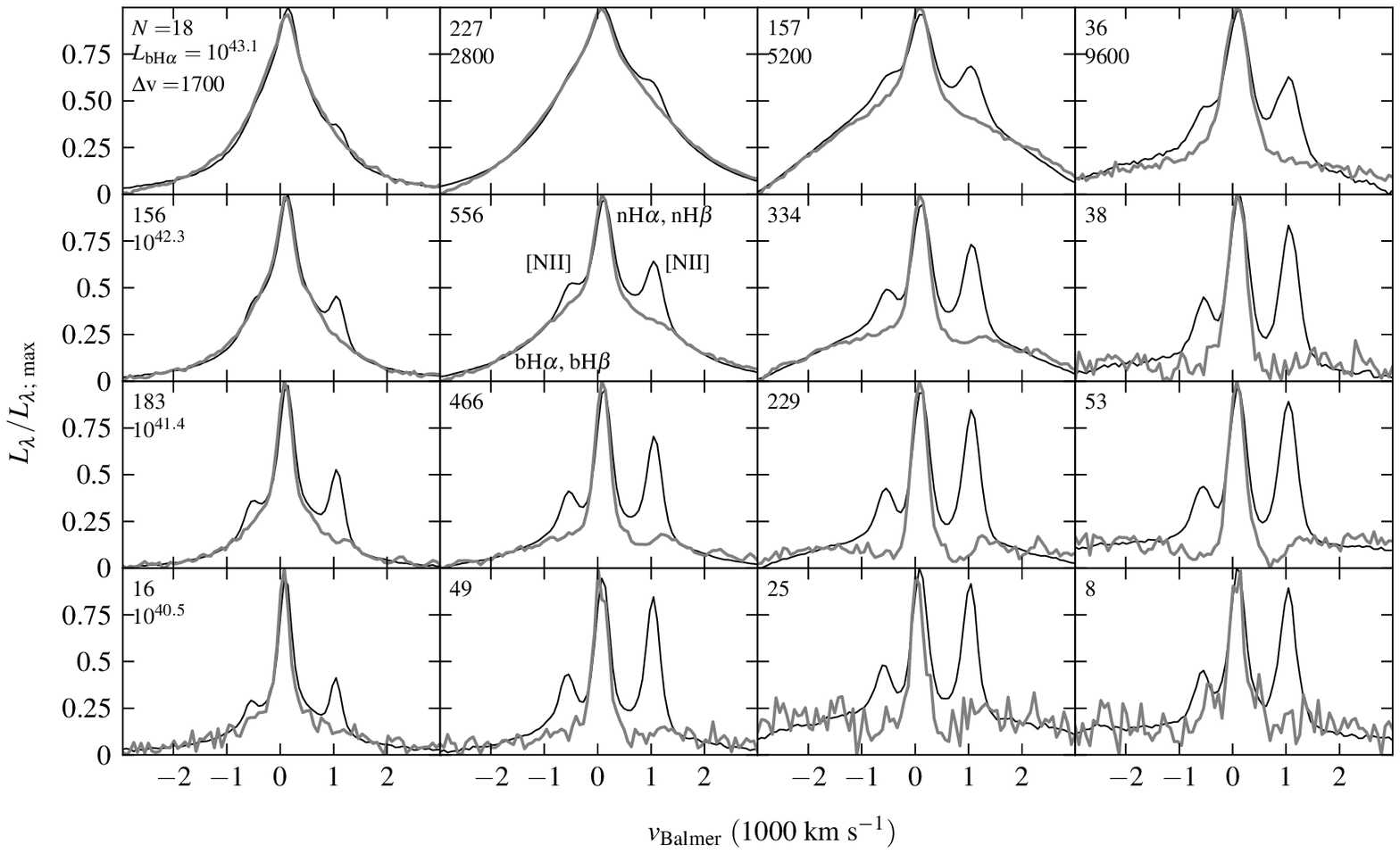}
\caption{
Mean host-subtracted spectra of the $\luv/\lbha<100$ T1 objects, near \Ha\ and near \Hb, at different $\lbha$ and $\dv$.
Each panel shows the mean spectrum of T1 objects with the same $\dv$ and the same $\lbha$ (number of objects noted). The $\lbha$ increases from bottom to top (mean $\lbha$ in $\ergs$ noted in left column), and $\dv$ increases from left to right (mean $\dv$ in $\kms$ noted in top row). 
The thin black line plots the mean spectrum centered around \Ha. 
To enhance the contrast between different NLR and BLR components, we also plot the mean spectrum centered around \Hb, with the $L_\lambda$ adjusted to fit the $L_\lambda$ of the \Ha\ region (thick gray line). 
In the $\dv = 2800,\ 5200,$ and $9600\ \kms$ columns, the $\nii/\Ha$ ratio clearly decreases with increasing $\lbha$.
}
\label{fig: }
\end{figure*}

In \S\S 3 -- 6, we show that the distribution of \nii/\Ha\ in the T1 sample shifts to lower values with increasing $\lbha$. 
To exclude the possibility that this trend is an artifact of our deblending algorithm, we examine the mean spectra at different $\lbha$ and $\dv$.

We divide the T1 objects with $\luv/\lbha<100$ (to avoid host contamination of the NLR) to bins of 0.3 dex in $\dv$ and one decade in $\lbha$. For each bin, we derive the host-subtracted mean spectrum, as described in \S6.1.2. These mean spectra are plotted in Figure A1, with $\lbha$ increasing from bottom to top, and $\dv$ increasing from left to right. The thin black line in each panel shows the relevant mean spectrum in the $L_\lambda$ vs. velocity $v$ plane, centered on \Ha. 
To enhance the contrast between different NLR and BLR components, we also plot the same mean spectrum centered on \Hb\ (thick gray line), with the $L_\lambda$ of the \Hb\ region fit to the $L_\lambda$ of the \Ha\ region. The fit is performed by a least square minimization of two parameters $a$ and $b$ so that 
\begin{equation}
 L_\lambda(v_\Ha) \approx a \times L_\lambda(v_\Hb) + b
\end{equation}
for all $v$ in the ranges $-3000 < v < -1000\ \kms$, $-200 < v < 450\ \kms$ and $1600 < v < 3000\ \kms$. The $v$ ranges for the fit are chosen to avoid the \nii\ lines. The best fit coefficient $a$ is between 2.7 and 4 in all panels.

The decrease in \nii/\Ha\ with increasing $\lbha$ is clear in the $\dv = 2800,\ 5200,$ and $9600\ \kms$ columns, confirming the trend found on single objects in \S\S 3--6.

\section{BPT by $\mbh$ and $M_*$}

Figures B1 and B2 show the BPT positions of the T1 sample, divided by $\mbh$ and $M_*$, in the same format as Figures 3--5 above. In Figure B2, only the 91\% of T1 objects with a reliable estimate of $M_*$ (\S2.3.1) are shown. 

\begin{figure*}
\includegraphics{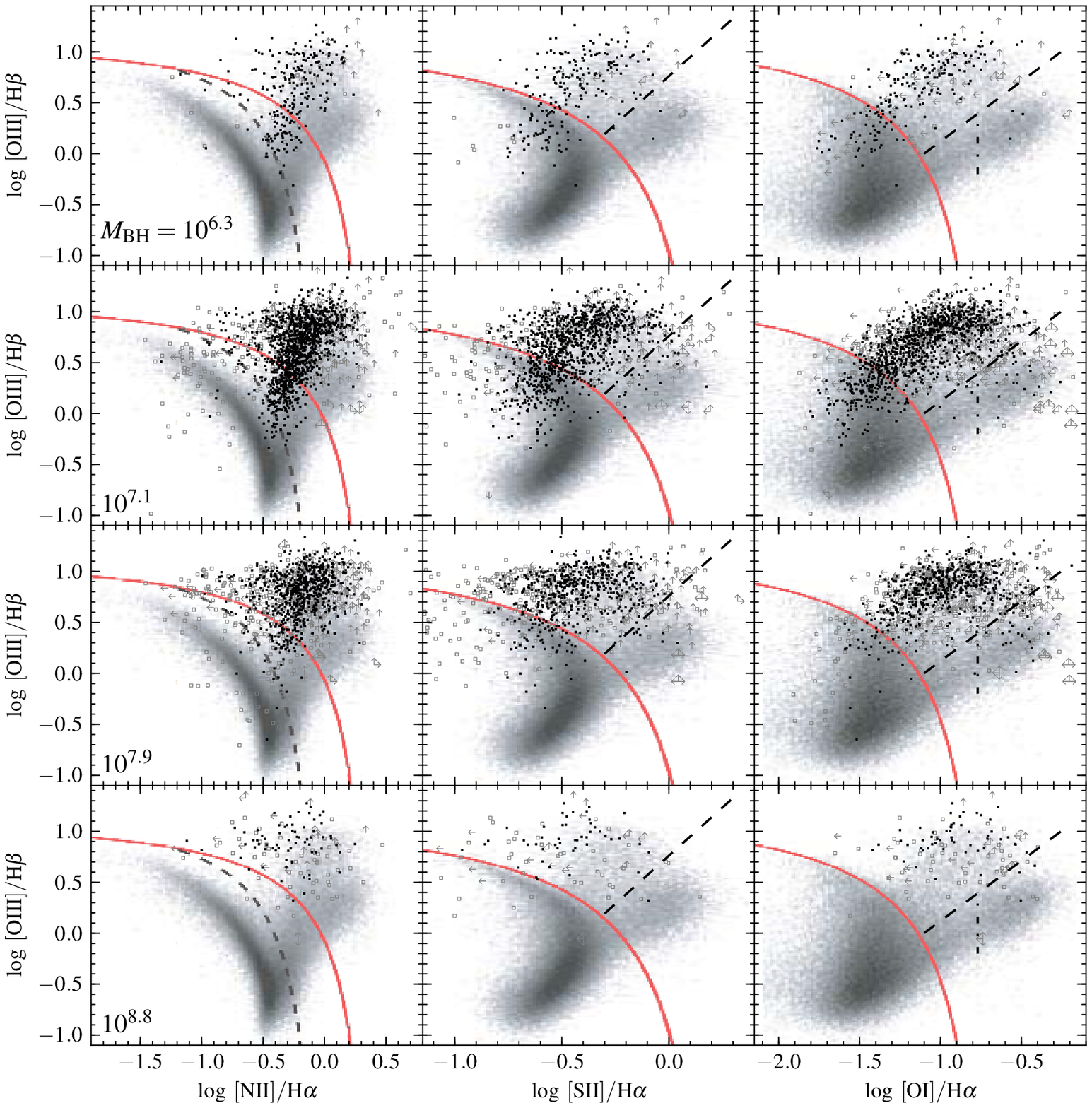}
\caption{
As in Figure 3, for the dependence of BPT position on $\mbh$. 
In each row, T1 AGN from a given decade-wide bin in $\mbh$ are plotted (mean $\mbh$ noted, in $\msun$).
The frequency of composites and SFs decreases with increasing $\mbh$, from 32\% at $\log\ \mbh = 6.3$, to 23\%, 14\% and 6\% at $\log\ \mbh = 7.1,\ 7.9$ and $8.8$, respectively.
}
\label{fig: }
\end{figure*}

\begin{figure*}
\includegraphics{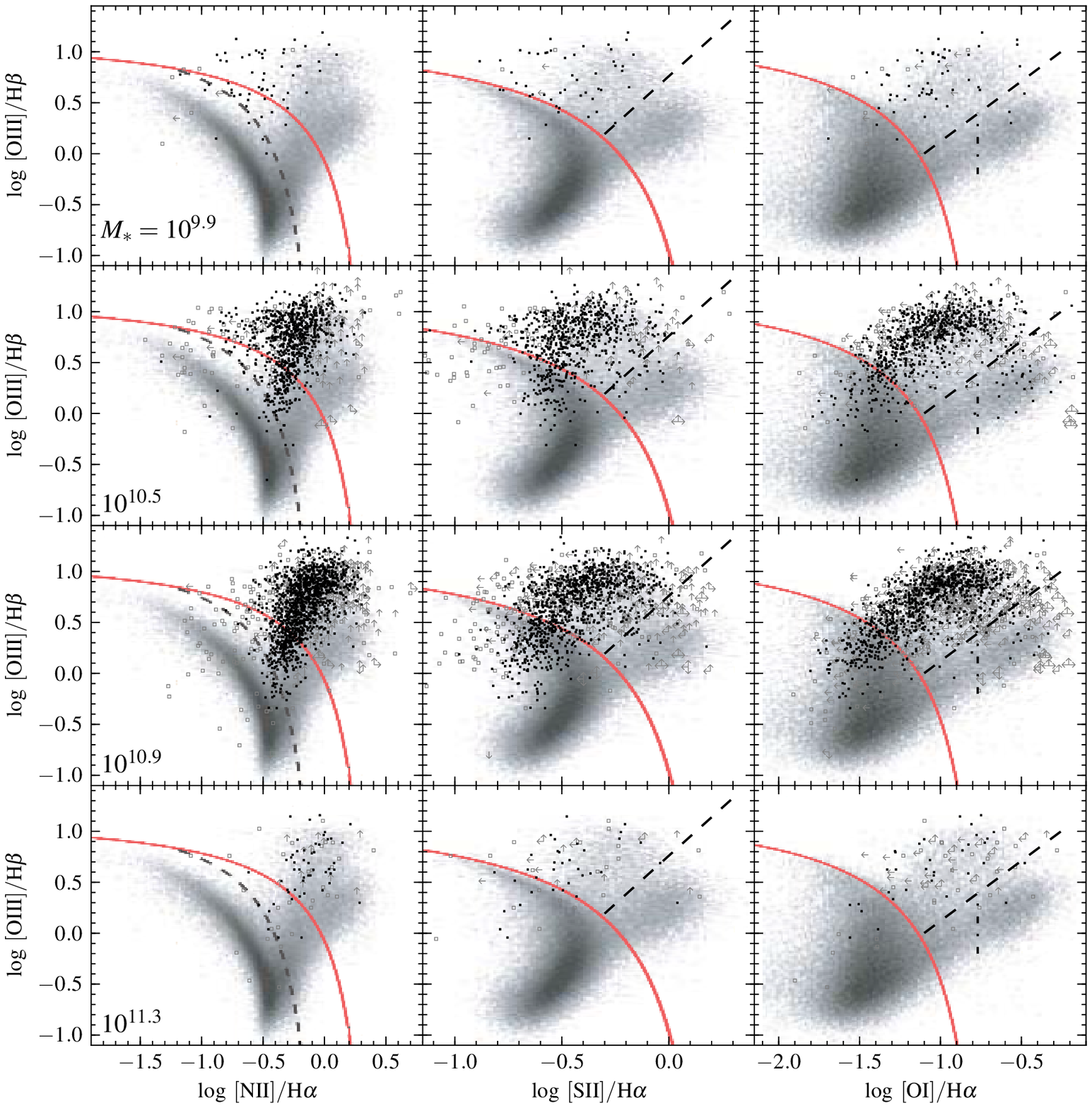}
\caption{
As in Figure 3, for the dependence of BPT position on $M_*$. 
In each row, T1 AGN from a given half decade wide bin in $M_*$ are plotted (mean $M_*$ noted, in $\msun$).
With decreasing $M_*$, an increasing fraction of objects are offset to low \nii/\Ha\ values, consistent with the $M_* - Z$ relation of quiescent galaxies.
}
\label{fig: }
\end{figure*}

With decreasing $\mbh$ the fraction of T1s classified as Composites and SFs increases, indicating an increase in the relative amount of host contribution to the NLR (\S8). Since the SDSS is a flux limited sample, T1s with low $\mbh$, and therefore low bulge mass, are preferentially selected from disk dominated galaxies (see Figure 16 in Paper I). Disks have a relatively large specific star formation rate, which may cause the observed shift in the BPT positions.

With decreasing $M_*$, an increasing fraction of objects are offset to low \nii/\Ha\ values, as found by Groves et al. (2006) on a type 2 AGN sample. This trend is consistent with the $M_* - Z$ relation of quiescent galaxies (Lequeux et al. 1979, and citations thereafter).

\section{The apparent Baldwin effect when using a proxy for $\lcont$}
\newcommand{\sxx}{\sigma_{\rm XX'}}
\newcommand{\syx}{\sigma_{\rm YX}}

Assume $Y$ and $X$ are some variables. If intrinsically $Y \propto X^{1.0}$, and one measures $Y$ and $X'$ on some sample, where $X'$ is a proxy for $X$, then due to the dispersion between $X$ and $X'$ one will find $Y \propto X'^{(1-\epsilon)}$. 
In \S3, $Y \equiv L_{\rm NL}$, $X \equiv \lcont$ and $X'\equiv \lbha$. Therefore, $\epsilon$ is the Baldwin effect one would measure when using $\lbha$ as a proxy for $\lcont$, assuming no intrinsic Baldwin effect. In this section, we evaluate $\epsilon$ analytically. 
We assume
\begin{equation}
  \begin{split}
   Y  &= 1 \cdot X + \syx  + b \\
   X' &= 1 \cdot X + \sxx  + c
  \end{split}
\end{equation}
where $b$ and $c$ are some constants, and $\sigma_{\rm AB}$ denotes the dispersion between $A$ and $B$. We assume the $\sxx$ and $\syx$ are independent of $X$ and of each other, and symmetric around zero. 
To significantly reduce the algebra, without affecting the final result, we set $b=c=\overline{X}=0$, where $\overline{X}$ is the mean $X$ in the sample.

The best fit slope is derived from:
\begin{equation}
 \frac{d}{d\epsilon} \frac{1}{N}\sum (Y - (1-\epsilon) X')^2 = 0
\end{equation}
Differentiating and dividing by $2$, the left side equals
\begin{equation}
 \begin{split}  
  & \frac{1}{N}\sum (Y - (1-\epsilon)X') X' = \\
  & \frac{1}{N}\sum YX' - (1-\epsilon)X'^2 = \\
  & \frac{1}{N}\sum (X + \syx)(X+\sxx) - (X+\sxx)^2 + \epsilon(X+\sxx)^2
 \end{split}
\end{equation}
Utilizing the assumptions on $\sxx$ and $\syx$ above, all terms which are linear in $\sxx$ or $\syx$ vanish for $N \rightarrow \infty$. Therefore, we are left with
\begin{equation}
 \frac{1}{N} \sum X^2 - X^2 - \sxx^2 + \epsilon(X^2 + \sxx^2)
\end{equation}
From $\overline{X}=0$ we get $\frac{1}{N} \sum X^2 = (\Delta X)^2$, where $\Delta X$ is the standard deviation of the distribution of $X$ spanned by the sample. We abuse notation a bit and replace $\frac{1}{N}\sum \sxx^2$ with $\sxx^2$. Therefore,
\begin{equation}
 -\sxx^2 + \epsilon( (\Delta X)^2 +\sxx^2 ) = 0
\end{equation}
and hence,
\begin{equation}
\begin{split}
  \epsilon  = \frac{\sxx^2}{(\Delta X)^2 +\sxx^2} 
             = \frac{1}{1 + \frac{(\Delta X)^2}{\sxx^2}} 
         \leq \frac{1}{1 + \frac{(\Delta X')^2 - \sxx^2}{\sxx^2} } 
       = (\frac{\sxx}{\Delta X'})^2
\end{split}
\end{equation}
As expected, $\epsilon$ decreases when the dynamical range of $X'$ increases. 

In \S3, we show that in the T1 sample eq. C6 implies $L_{\rm NL} \propto \lbha^{1-0.07}$. The observed slopes of $\lesssim 0.7$ are significantly lower than 0.93, and therefore imply the existence of an intrinsic Baldwin effect.

\label{lastpage}

\end{document}